\documentclass[%
 reprint,
 amsmath,amssymb,
 aps,
]{revtex4-1}
\usepackage{color}
\usepackage{graphicx}
\usepackage{dcolumn}
\usepackage{bm}
\RequirePackage{color}

\RequirePackage[colorlinks,citecolor=blue,urlcolor=blue,linkcolor=blue]{hyperref}

\begin{document}

\preprint{APS/123-QED}

\title{Curved non-interacting two-dimensional electron gas with anisotropic mass}

\author{Pedro H. Souza}
\affiliation{Departamento de F\'{\i}sica,
           Universidade Federal de Lavras,
           Caixa Postal 3037,
            37200-000 Lavras-MG, Brazil} 
\author{Edilberto O. Silva}
\email{edilberto.silva@df.ufma.br}
\affiliation{Departamento de F\'{\i}sica, Universidade Federal do Maranh\~{a}o, 65085-580 S\~{a}o Lu\'{\i}s-MA, Brazil}
\author{Moises Rojas}
\email{moises.leyva@dfi.ufla.br}
\affiliation{Departamento de F\'{\i}sica,
           Universidade Federal de Lavras,
           Caixa Postal 3037,
            37200-000 Lavras-MG, Brazil}
\author{Cleverson Filgueiras}
\email{cleverson.filgueiras@dfi.ufla.br}
\affiliation{Departamento de F\'{\i}sica,
           Universidade Federal de Lavras,
           Caixa Postal 3037,
            37200-000 Lavras-MG, Brazil}

\date{\today}

\begin{abstract}
In the da Costa's thin-layer approach, a quantum particle moving in a 3D sample is confined on a curved thin interface. At the end, the interface effects are ignored and such quantum particle is localized on a curved surface. A geometric potential arises and, since it manifests due to this confinement procedure, it depends on the transverse to the surface mass component. This inspired us to consider, in this paper, the effects due to an anisotropic effective mass on a non-interacting two dimensional electron gas confined on a curved surface, a fact not explored before in this context. By tailoring the mass, many investigations carried out in the literature can be improved which in turns can be useful to better designing electronic systems without modifying the geometry of a given system. Some examples are examined here, as a particle on helicoidal surface, on a cylinder, on a catenoid and on a cone, with some possible applications briefly discussed.
\begin{description}
\item[PACS numbers]
 03.65.Ca, 02.40.-k, 68.65.-k
\end{description}
\end{abstract}

\maketitle

\section{Introduction}
Due to the technological progress, like the one reported about the synthesis of Si and Ge nanocones with controllable apex angles \cite{coneexperiment}, the quantization of constrained motions for a nonrelativistic particle in lower-dimensional systems and nanostructures has being receiving considered attention over the years. The ability of building two dimensional curved substrates in desired shapes corroborate with this growing interest \cite{shapes}. In Ref. \cite{cylinderexperiment}, structures with a high-mobility two-dimensional electron gas (2DEG) constrained on a cylindrical surface were fabricated and a giant asymmetry in the measured  magnetoresistance was revealed. As an interesting theoretical work, we may refer to the Ref. \cite{introducao3}, where it was showed that a core-shell nanowire can be modeled by a cylindrical surface of finite length and a calculation of the conductance by coupling it to leads in the presence of a uniform magnetic field perpendicular to the axis of the cylinder leaves to the existence of the snaking states which govern transport at low chemical potential. In \cite{PhysRevLett.113.046803}, a general method to compute correlation functions of fractional quantum Hall states on a curved surface was developed.

In the continuum, the start point to investigate a non-interacting curved 2DEG is the da Costa's approach \cite{dacosta}, where the Schr\"{o}dinger equation of a free quantum particle constrained to move in an infinitely thin curved interface of the ordinary three-dimensional space, was derived. The  splits into the normal and a tangent part.  After the separation of these modes, a geometric quantum potential given in terms of both the Gaussian and the mean curvatures arises. Among the vast applications which followed such a work, we have the case of a non-interacting 2DEG constrained to move in a helicoid. In this geometry, by considering a helicoidal nanoribbon, a reminiscent of the Hall effect induced by the geometric quantum potential \cite{PhysRevLett.100.230403} in the nonrelativistic limit was pointed out,  a fact which was also observed in the case of carriers in graphene shaped in the same way \cite{PhysRevB.92.035440}.

The version for a curved non-interacting 2DEG in the presence of an electric and magnetic field was investigated in \cite{PhysRevLett.100.230403}, where it was showed that no coupling between the fields and the quantum geometric potential arises. An application in this case can be viewed in \cite{conePRL}, where important features concerned to the quantum Hall states on surfaces with conical singularities has been presented. The version for Pauli equation for a charged spin particle on a curved surface in an electric and magnetic field has been addressed in \cite{paulicurved}. An application which followed this reference can be viewed in \cite{truncado}. The refinement of the fundamental framework of the thin-layer quantization procedure \cite{dacosta,PhysRevLett.100.230403} considering the surface thickness has been carried out in \cite{Wang201668}. We will not consider such  influence of the surface thickness here since the extra terms in the geometric potential including the influence of the surface thickness must be treated using the perturbation theory. More important recent theoretical works in this field of curved 2DEG can be viewed in \cite{Xun2014132,Spittel:15,Jahangiri2016407,Panahi201657,Wang2016,Cruz2017,centripetal}. Based on what has been exposed here so far, we conclude that curvature is an important degree of freedom to manipulate two dimensional quantum systems. Most of the applications of the thin-layer quantization are concerned to common materials, like carbon nanotubes, Si, Ge, GaAs, etc. This means that the carrier {\it effective mass} has an important role in those problems. From the band theory of solids, such effective mass means that the inertia of particles in a periodic potential, over long distances larger than the lattice spacing, is not the same as its motion in a vacuum \cite{kittel2005introduction}. The movement of a particle  in the crystal is modeled as it was a free particle with such mass. Depending on the purpose, it can be considered as a simple constant of a given material. It is an important parameter to tailor  quantum systems \cite{tailoring,silverinha}. In general, its value depends on the purpose for which it is used, depending on a number of other parameters.

In this paper, we consider the effects due to an anisotropic effective mass on a non-interacting 2DEG confined on a curved surface. We follow the da Costa's approach mentioned above. The curvature potential is attractive for non-minimal surfaces in the case of isotropic mass. By tailoring the mass, this scenario can be changed, allowing us to have different physical phenomena for electronic systems without modifying the geometry of the surface.

This paper is organized as follows: In section \ref{sec2}, the considerations about the anisotropic mass are presented. In section \ref{sec3}, the da Costa's approach is briefly reviewed but now taking into account such anisotropic mass. Some applications are considered in section \ref{sec4}. We have the concluding remarks in section \ref{sec5}.

\section{The anisotropic effective mass}\label{sec2}
Considering the continuum limit, the energy levels of spinless electrons in semiconductors are obtained from the effective mass Schr\"{o}dinger equation,
\begin{equation}
    -\frac{\hbar^2}{2}\left[\frac{1}{m^*}\right]^{ij}\partial_i\partial_j \Psi+V\Psi=E\Psi, \label{ef-mass-schro}
\end{equation}
where the effective mass tensor $[1/m^*]^{ij}$ in a diagonalized form is given by \cite{anisotropicmass}
\begin{equation}
\left[\frac{1}{m^*}\right]^{ij}=\left(\frac{1}{\hbar^2}\frac{\partial^2 E}{\partial k_i\partial k_j}\right)_{\mathbf{k}=0}=\begin{pmatrix}
    m_{11}^{-1}     & 0             & 0 \\
    0               & m_{22}^{-1}   & 0 \\
    0               & 0             & m_{33}^{-1}
\end{pmatrix}.
\label{emass}
\end{equation}
The crucial point in this paper is that the principal masses $m_{11}$, $m_{22}$ and $m_{33}$ can be not only equal but also different form each other, including the case of negative effective mass as well \cite{dragoman2007metamaterials}. In it, an electronic metamaterial \cite{metareview,JPCOmeta} can be modeled with the effective mass $m^*$ and the difference $(E-V)$ being the electronic counterparts of the electric permittivity $\epsilon$ and the magnetic permeability $\mu$, respectively. This way, a positive (negative) effective mass corresponds to a positive (negative) permittivity. The same is true for $(E-V)$ and $\mu$. We will consider the electronic analogue of a hyperbolic metamaterial with permittivity tensor
\begin{equation}
    \epsilon_{ij}=\begin{pmatrix}
        \epsilon_1 & 0 & 0\\
        0 & \epsilon_1 & 0\\
        0 & 0 & -|\epsilon_2|
\end{pmatrix},
\label{epsilon_tensor}
\end{equation}
with $\epsilon_{xx}=\epsilon_{yy}=\epsilon_1>0,\epsilon_{zz}=\epsilon_2<0$. The following analog effective mass tensor is
\begin{equation}
    \left[\frac{1}{m^*}\right]^{ij}=\begin{pmatrix}
        m_1^{-1} & 0 & 0\\
        0 & m_1^{-1} & 0\\
        0 & 0 & -|m_2|^{-1}
\end{pmatrix},
\label{m_tensor}
\end{equation}
where $m_{11}=m_{22}=m_1>0,m_{33}=m_2<0$.
It is interesting to point out that such analogy has led to a physical link between the metamaterial and the propagation of Klein-Gordon particles in flat background spacetime exhibiting discontinuous metric changes from a Lorentzian signature $(-,+,+,+)$ to a Kleinian
signature \cite{PhysRevD.94.044039}.
The difference existent between these cases is that the dispersion relation of a metamaterial with permittivity tensor (\ref{epsilon_tensor}) is
a hyperboloid of two sheets in $\mathbf{k}$ space \cite{shekhar2014hyperbolic}, while that the dispersion relation for the matter waves is an ellipsoid in $\mathbf{k}$ space.
\section{Schr\"{o}dinger equation for a curved non-interacting 2DEG with anisotropic mass}\label{sec3}
In this section we will derive the Schr\"{o}dinger equation valid for any 2D geometry, describing curved nanostructures considering the effective mass tensor given by (\ref{emass}) with $m_{11}=m_{22}\equiv m_1$ and $m_{33}\equiv m_2$.
In what follows, $i,j$ stand for the spatial indexes with the values $1,2,3$. Covariant and contravariant Tensor components are used and Einstein summation convention is adopted. The covariant 3D Schr\"{o}dinger equation in a generic 3D curvilinear coordinate system in the absence of electric and magnetic fields is
\begin{equation}
\label{eq:schr3}
i\hbar\partial_t \psi=
-\frac{\hbar^2}{2}\left[\frac{1}{m^*}\right]^{i^{\prime}j^{\prime}}
\left[
\frac{1}{\sqrt{G}}\partial_i\left(\sqrt{G} G^{ij}\partial_j\right)\psi\right],
\end{equation}
where $G_{ij}$ is the metric tensor, $G^{ij}$ is its inverse, $G=\det(G_{ij})$ and $i^{\prime},j^{\prime}=1,2,3$.

Following the thin-layer procedure described by da Costa \cite{dacosta} to confine the particle on the surface, we consider the surface $S$ parametrized by $\vec{r}=\vec{r}(q_1,q_2)$,where $\vec{r}$ is the position vector of an arbitrary point on the surface.
The 3D space in the immediate neighborhood of $S$ can be parameterized as
$\vec{R}(q_1,q_2,q_3)=\vec{r}(q_1,q_2)+q_3\hat{n}(q_1,q_2)$,
where $\hat{n}(q_1,q_2)$ is a unit vector normal to $S$.
The indexes $a,b$ stand for the surface parameters, which assume the values $1,2$.
The relation between the 3D metric tensor $G_{ij}$ and the 2D induced one, $g_{ab}={\partial_a \vec{r}}\cdot{\partial_b \vec{r}}$, is given by
\begin{eqnarray}
\label{eq:metric}
G_{ab}=g_{ab}+\left[\alpha g+(\alpha g)^T\right]_{ab}q_3+(\alpha g \alpha^T)_{ab}q_3^2\nonumber\\
\\
G_{a3}=G_{3a}=0, \; G_{33}=1\nonumber,
\end{eqnarray}
where $\alpha_{ab}$ is the Weingarten curvature matrix for the surface \cite{dacosta,Wolfram}.
The structure of the metric tensor given in (\ref{eq:metric}) suggests to separate the Schr\"{o}dinger (\ref{eq:schr3}) in a surface part for $a,b=1,2$ and a normal part, for $i,j=3$. In such well-established thin-layer method \cite{dacosta}, a confining potential $V_{\lambda}(q_3)$ is considered in order to localize the particle on the surface $S$, where $\lambda$ is a parameter which measures the strength of the confinement. The aim of the procedure is to obtain a surface  which depends only on $(q_1,q_2)$. This way, a new wave function
$
\chi(q_1,q_2,q_3)=\chi_{S}(q_1,q_2) \chi_{n}(q_3)\;\label{new}
$
is introduced. The condition of conservation of the norm yields the relation
\begin{equation}
\label{eq:chichi}
\psi(q_1,q_2,q_3)=\left[ 1+{\rm Tr}(\alpha)q_3+\det(\alpha)q_3^2 \right]^{-\frac{1}{2}}\chi(q_1,q_2,q_3)\;.
\end{equation}
As demonstrated in \cite{dacosta}, the structure of the metric tensor given by (\ref{eq:metric}), allows the Laplacian to be broken into two parts: the surface part, denoted by $\triangle$, given by the terms $a,b=1,2$, and the normal part, defined by $i,j=3$. Unlike \cite{dacosta}, we consider the mass tensor in a way that the surface effective mass will be given by $m_1$, while that the normal one is $m_2$, with $m_1\neq m_2$. Considering Eq. (\ref{eq:schr3}) in this case, the Schr\"{o}dinger equation can be written as
\begin{eqnarray}
-\frac{\hbar^{2}}{2m_1}\triangle\psi&-&\frac{\hbar^{2}}{2m_2}\left(\frac{\partial^{2}}{\partial q_{3}^{2}}+\frac{\partial\left(\ln\sqrt{G}\right)}{\partial q_{3}}\frac{\partial}{\partial q_{3}}\right)\psi\nonumber\\ &+&V_{\lambda}\left(q_{3}\right)\psi=i\hbar\frac{\partial}{\partial t}\psi\;.\label{new}
\end{eqnarray}
Substituting (\ref{eq:chichi}) into (\ref{new}), the wave function will be localized on $S$ in the limit of confinement which consist of two step potential barriers on both sides of the surface.
Performing the limit $q_3\to0$ in the Schr\"{o}dinger equation, it yields
\begin{eqnarray}
\label{eq:shro2D}
i\hbar \partial_t \chi&=&
-\frac{\hbar^2}{2m_1}
\left[
\frac{1}{\sqrt{g}}\partial_a\left(\sqrt{g}g^{ab}\partial_b \chi\right)\right]+
V_{S}(q_1,q_2)\chi
\nonumber\\ &-&\frac{\hbar^2}{
2m_2}\left(\partial_3\right)^2\chi+V_{\lambda}(q_3)\chi,\nonumber
\end{eqnarray}
where $g=\det{(g_{ab})}$ and $V_{S}$ is
the well-known geometric potential, given by  \cite{dacosta}
\begin{equation}
V_{S}(q_1,q_2)=-\frac{\hbar^2}{2m_2}\left(\left[\frac{1}{2}{\rm Tr}(\alpha)\right]^2-\det(\alpha)\right),
\end{equation}
On it, the first term is the square of the mean curvature,
$\rm M=\rm Tr(\alpha)/2=(\kappa_1+\kappa_2)/2$ and the second one is the Gaussian curvature,
${\rm K_G}=\det(\alpha)=\kappa_1 \kappa_2$. $\kappa_1$ and $\kappa_2$ are the principal curvatures at a given point of the surface. Notice that the curvature potential came from the normal part of the Schr\"{o}dinger equation before separation. Although we consider the limit $q_3\rightarrow0$, we have in fact a quasi-2DEG, meaning that the transverse component of the mass will have some impact in the dynamics of the electrons, since the geometric potential arises by considering that the curved surface is immersed in a 3D space. If we consider the problem strictly two dimensional, without the confinement procedure, such geometric potential does not arises.

In the case for $m_2<0$, the geometry induced potential reverses its sign. It seems simple the fact that considering the transverse component of the mass tensor as negative, the only modification is the sign of $V_S$. But the impact of such modification is rouge, since we go from an attractive geometric quantum potential in ordinary material to a repulsive one in a metamaterial, when the surface is not a minimal one (${\rm M}\neq0$). If we have a minimal surface, then ${\rm M}\equiv 0$ and the geometry quantum potential change its profile: in regions where it is attractive, it becomes repulsive, and vice-versa. We will examine some examples below to show how important these modifications are.

Next, it is defined a new metric tensor
\begin{equation}
\tilde{G}=
\left(
\begin{array}{ccc}
g_{11}&g_{12}&0\\
g_{21}&g_{22}&0\\
0&0&1
\end{array}
\right),
\end{equation}
Finally, the separability of the dynamics on the surface and perpendicular to the surface leaves to the two independent Schr\"{o}dinger equations,
\begin{eqnarray}
\hbar \partial_t \chi_{n}=-\frac{\hbar^2}{2m_2}(\partial_3)^2\chi_{n}+V_{\lambda}(q_3)\chi_{n},\label{eq:dynq3}
\end{eqnarray}
\begin{eqnarray}
\hbar \partial_t \chi_{S}&=&
-\frac{\hbar^2}{2m_1}
\left[
\frac{1}{\sqrt{g}}\partial_a\left(\sqrt{g}g^{ab}\partial_b \chi_{S}\right)
\right]\nonumber\\
&-&\frac{\hbar^2}{2m_2}\left[\left(\frac{1}{2}{\rm M}\right)^2-{\rm K_G}\right]\chi_{S}\;.
\label{eq:shrodecop}
\end{eqnarray}
In Ref. \cite{PhysRevLett.100.230403}, it was showed that, in the presence of magnetic and electric fields, there is no coupling between these fields and the surface curvature. Therefore, the surface and transverse dynamics are exactly separable as above, but with (\ref{eq:shrodecop}) written as
\begin{align}
i\hbar \partial_{t}\chi _{S}& =\frac{1}{2m_{1}}\Big\{-\frac{\hbar ^{2}}{\sqrt{g}%
}\partial_{a}\left(\sqrt{g}g^{ab}\partial _{b}\chi _{S}\right)  \notag\\
& +\frac{iQ\hbar}{\sqrt{g}}\partial_{a}\left(\sqrt{g}g^{ab}A_{b}\right)
\chi _{S}+2iQ\hbar g^{ab}A_{a}\partial _{b}\chi_{S} \notag\\
& +Q^{2}g^{ab}A_{a}A_{b}\chi _{S}\Big\}-\frac{\hbar ^{2}}{2m_{2}}\left[ \left(
\frac{1}{2}\mathrm{M}\right) ^{2}-\mathrm{K_{G}}\right] \chi _{S} \notag\\
& +QV\chi _{S}\;, \label{eq:shrodecop2}
\end{align}
where $Q$ is the electrical charge of a quantum particle and $A_{a}$ are the covariant components of the potential vector $\vec{A}$.

In what follows, we consider some anisotropic values of the effective mass based on references \cite{negativemass,silverinha,tailoring}.
\section{Some applications}\label{sec4}
\subsection{A quantum particle on a helicoid}
\begin{figure}
\includegraphics[width=0.4\textwidth]{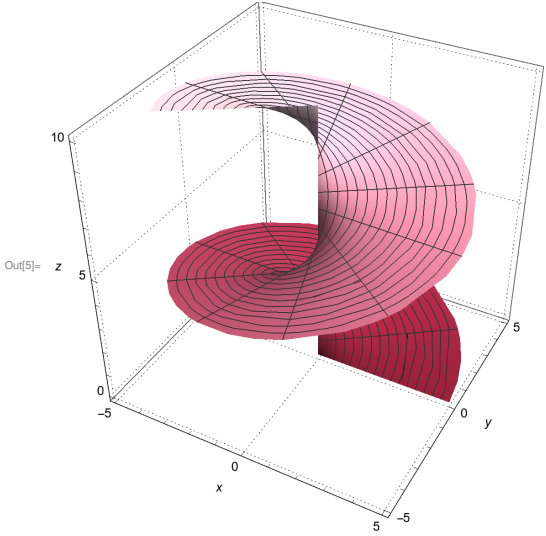}
\caption{A helicoid} \label{helicoidpng}
\end{figure}

A helicoid, showed in Fig. \ref{helicoidpng}, can be parametrized by the following set of equations \cite{geometry}:
\begin{eqnarray}
x &=& \rho\cos(\omega z), \label{E:xparaheli} \nonumber \\
y &=& \rho\sin(\omega z), \label{E:yparaheli} \nonumber \\
z &=& z, \label{E:zparaheli}
\end{eqnarray}
where $\omega=2\pi S$, with $S$ being the number of complete twists (i.e., $2\pi$-turns) per unit length of the helicoid. $\rho$ is the radial distance from the $z$-axis.  The infinitesimal line element on the helicoid is given by
\begin{equation} \label{E:lineelementheli}
ds^2 = d\rho^2 + (1 + \omega^2\rho^2)dz^2\;,
\end{equation}
where we have used the coordinates $(z, \rho)$ to characterize a point on the helicoid. The metric components are thus obtained as
\begin{eqnarray}
g_{\rho \rho} &=& 1, \nonumber \\
g_{zz} &=& 1 + \omega^2\rho^2, \nonumber \\
g_{z \rho} &=& 0, \nonumber \\
g_{\rho z} &=& 0\;. \label{E:metriczheli}
\end{eqnarray}
The square root of the determinant of the metric is given by
\begin{equation} \label{E:metricdeterminantheli}
\sqrt{g} = \sqrt{1 + \omega^2\rho^2}.
\end{equation}
The principal curvatures, $\kappa_1$ and $\kappa_2$, are given by
\begin{eqnarray}
\kappa_1 &=& \frac{\omega}{1+\omega^2\rho^2}, \nonumber \\
\kappa_2 &=& -\frac{\omega}{1+\omega^2\rho^2}\;. \label{helic}
\end{eqnarray}

A helicoid is a minimal surface, which means that the mean curvature ${\rm M}$ vanishes at any given point on it, that is,
\begin{equation} \label{E:meancurvatureheli}
{\rm M} \equiv \frac{1}{2}(\kappa_1 + \kappa_2) = 0\;.
\end{equation}
The Gaussian curvature ${\rm K_G}$ is
\begin{equation} \label{E:gaussiancurvatureheli}
{\rm K_G} = \kappa_1\kappa_2 = -\frac{\omega^2}{(1+\omega^2\rho^2)^2}\;.
\end{equation}
Therefore the geometric quantum potential $V_S$ will be read as
\begin{equation} \label{E:curvpotentialheli}
V_S = -\frac{\hbar^2}{2m_2}\left({\rm M}^2 - {\rm K_G} \right) =  -\frac{\hbar^2}{2m_2}\frac{\omega^2}{(1+\omega^2\rho^2)^2}\;.
\end{equation}
From (\ref{eq:shrodecop}), the Schr\"{o}dinger equation for a particle on a helicoid will be given by
\begin{align} \label{E:hamiltoniancurv}
i\hbar \partial_t \chi_{S}&=-\frac{\hbar^2}{2m_1}\left[\frac{1}{a} \left(  \partial_z( \frac{1}{a}\partial_z\chi_S ) + \partial_{\rho}(a\partial_{\rho}\chi_S )             \right)  \right] \notag \\
&- \frac{\hbar^2}{2m_2}\frac{\omega^2}{(1+\omega^2\rho^2)^2}\chi_s\;,
\end{align}
with $a\equiv \sqrt{1 + \omega^2\rho^2}$.
Considering the ansatz as
\begin{equation}
\chi_{S}=\exp\left(il\omega z\right)f\left(\rho\right)\;,
 \end{equation}
with $l\epsilon \mathbb{N}$ and considering that the wave function has to be normalized with respect to the infinitesimal area $d\rho dz$, we make the substitution $\chi_S \to \frac{1}{\sqrt{a}}\chi_S$ in Eq. (\ref{E:hamiltoniancurv}), which would only affect terms involving derivatives with respect to $\rho$ \cite{atanasov}. Omitting the algebra, we rewrite Eq. (\ref{E:hamiltoniancurv}) as
\begin{eqnarray} \label{E:hamiltoniancurvnorm}
H_{curv}\chi_{s}=-\frac{\hbar^{2}}{2m_{1}}\frac{d^{2}\chi_{s}}{d\rho^{2}}+\frac{\hbar^{2}}{2m_{1}}\left[\frac{l^{2}\omega^{2}}{1+\omega^{2}\rho^{2}}\right.-\nonumber\\ \left.\frac{\omega^{2}}{2(1+\omega^{2}\rho^{2})^{2}}\left(-1+\frac{\omega^{2}\rho^{2}}{2}+\frac{2m_{1}}{m_{2}}\right)\right]\chi_{s}\;.
\end{eqnarray}
For $m_1=m_2>0$, the result found in \cite{atanasov} is recovered.

\begin{figure}[!htp]
\centering
   \includegraphics[width=0.7\linewidth]{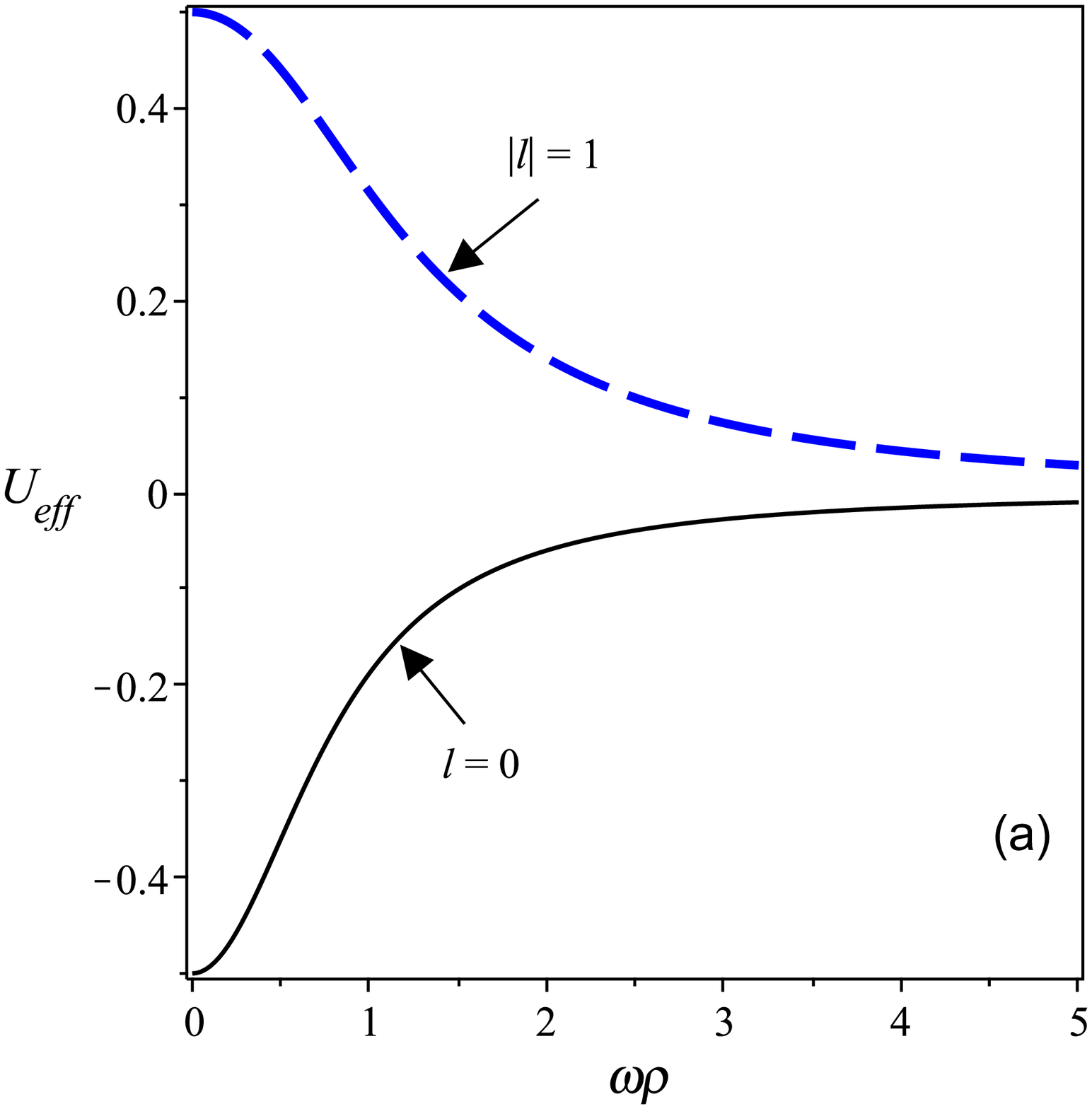}

   \includegraphics[width=0.7\linewidth]{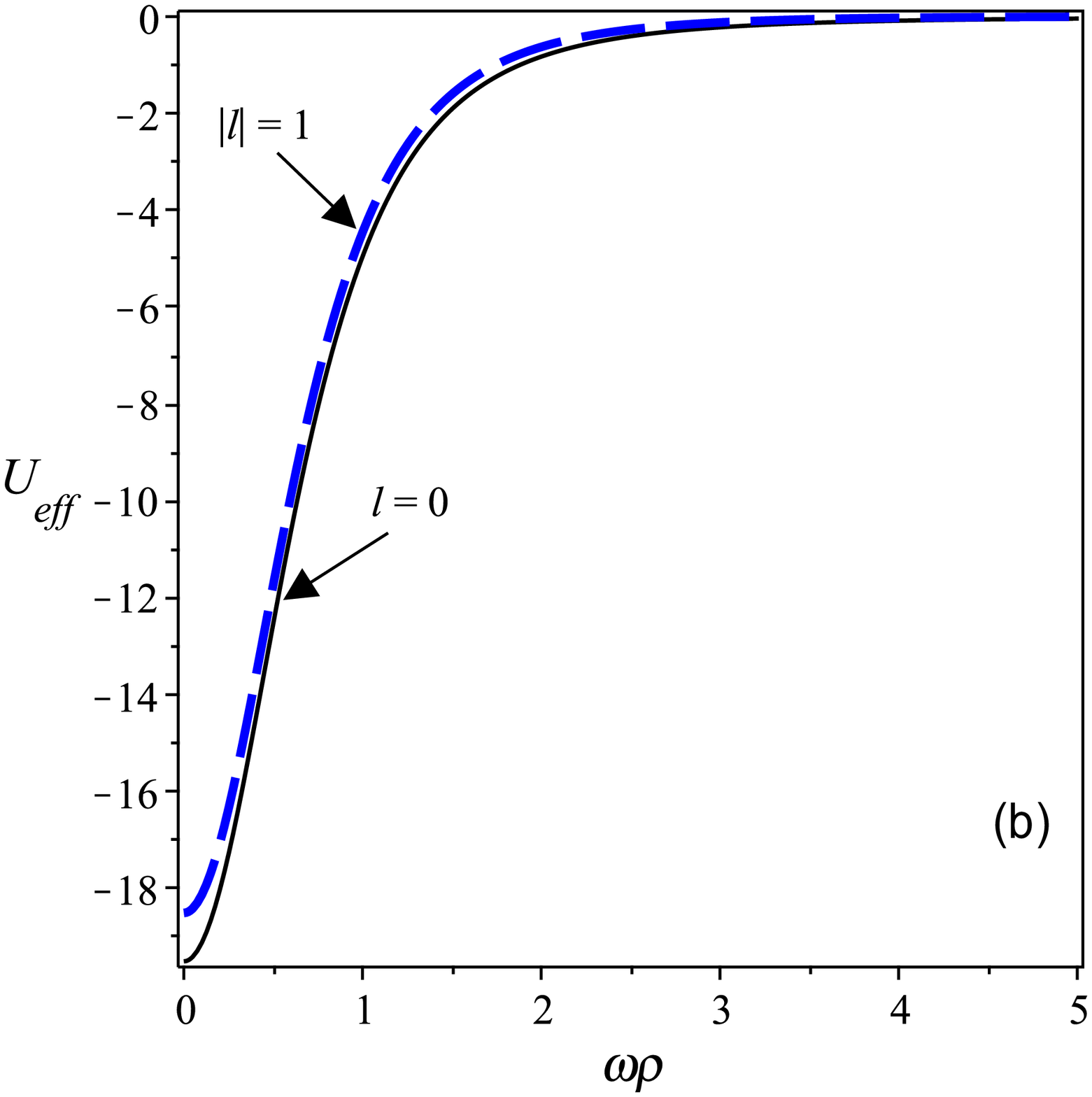}

   \includegraphics[width=0.7\linewidth]{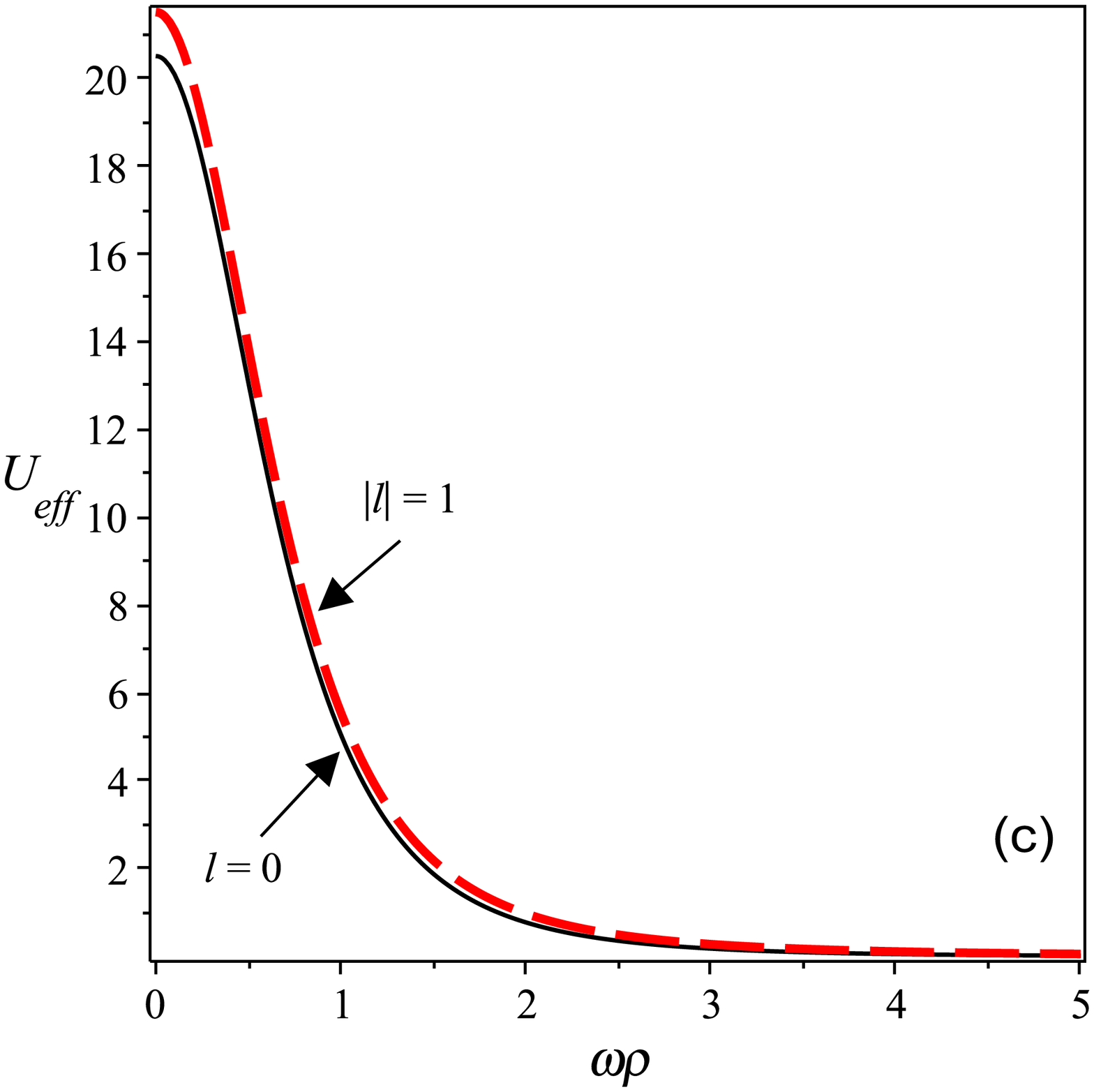}
\caption{\small The behavior of the curvature induced potential on a helicoid. In {\bf (a)}, we have an isotropic material ($m_1=m_2$), in {\bf (b)}, we have $m_1\neq m_2>0$ (quantum well) and an hyperbolic metamaterial in {\bf (c)} ($m_1>0$ and $m_2<0$, quantum barrier).}
\label{helichall}
\end{figure}

In Fig. \ref{helichall}, it is plotted the effective induced potential due to the helicoid for an isotropic material, an anisotropic one and for a hyperbolic metamaterial.  For the former,  the twist $\omega$ will  push  the  electrons  with $l\neq0$ ($l=0$) towards  the  outer  (inner)  edge  of  the  ribbon  and create an effective electric field between the central axis and the  helix,  the  latter  representing  the  rim  of  the  helicoid \cite{atanasov}.

Noticed that the charge separation due to the geometric quantum potential does not occur when $m_1\neq m_2>0$. In this case, for an infinity helicoid, all the electrons are repelled towards $\rho=0$ meaning that they can localize around $z-$axis. If we build a finite helicoid ($0<z<b$ and $0\leq\rho\leq a$), we have a quantum well which can be thought in the context of a quantum dot like structure,  where we may have bound states around the core of the helicoid and the phenomena concerned about optical absorption/emission can be explored in this system. The ratio between the masses and the pitch $\omega$ of the helicoid can be adjusted in order to manipulate these light absorption/emission. Moreover, studies concerned about optical rectification \cite{rectification}, second and third harmonic generations \cite{secthird} could be carried out as well.  In the case of a hyperbolic metamaterial ($m_2<0$), a quantum barrier is formed and all the electrons are repelled from the core of the helicoid.
\subsection{Transmission properties on isotropic/anisotropic cylindrical junctions}

\begin{figure}
\includegraphics[width=0.4\textwidth]{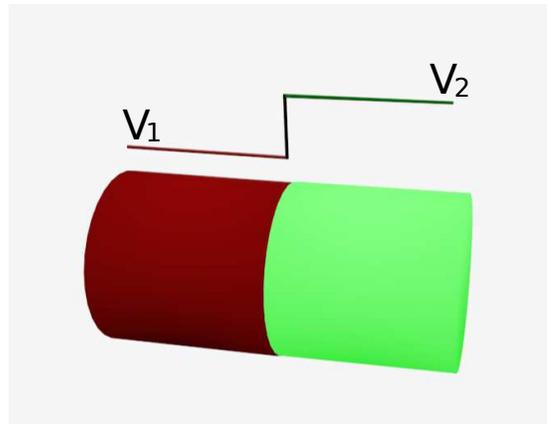}
\caption{A cylinder showing the two portions, one representing an ordinary material and the other one representing the metamaterial.} \label{cylinder}
\end{figure}

In Ref. \cite{bentcylinder}, the authors have showed that, by smoothly deforming a cylinder yielding a cylindrical junction, the effective potential originated by such geometry are going to be step-like quantum potentials. Then, a coherent-transport can be studied by considering the transmission properties of the system. In fact, the transmission and reflexion coefficients as a function of the geometrical parameters where obtained by them. Here, such transmission properties can be obtained by simple connecting two cylinders, one as a isotropic material an the other one as an anisotropic material (Fig. \ref{cylinder}). In this case,  rectangular quantum potential may arise, which are modeled as a Heaviside step function. Obviously, this case can be combined with those investigate in Ref. \cite{bentcylinder} in other to better manipulate and to control the quantum transmission properties in curved cylindrical junctions. Here, we will consider only the step quantum potential case, with no deformations in the cylindrical geometry.

\begin{figure}
\includegraphics[width=0.4\textwidth]{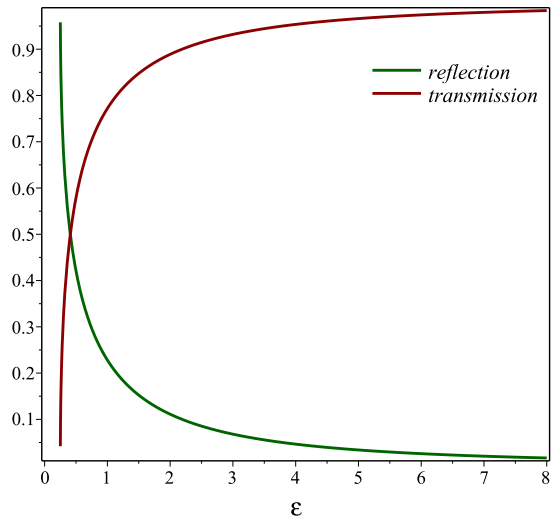}
\caption{Transmission and reflection coefficients for $L_\phi\equiv0$ versus $\varepsilon\equiv (2m_1R^2E)/\hbar^{2}$.}\label{rt}
\end{figure}

Consider a cylinder described by the line element
\begin{eqnarray}
dl^{2}=dz^{2}+R^{2}d\phi^{2}\;,
\end{eqnarray}
where $-\infty <z<\infty$ and $0\leq\phi<2\pi$. We divide the cylinder in two portions: red, for $z<0$ (isotropic material) and green for $z>0$ (anisotropic material). For simplicity, we consider the longitudinal components of the mass as the same in both regions. The transverse mass is $m_2\equiv m_1$, for $z<0$ and $m_2^{\prime}\equiv m_2$ for $z>0$. This way, the geometry induce potential in both regions are \begin{align}
V_1&=-\frac{\hbar^{2}}{8m_1 R^{2}}, \;{\text {for}}\; z<0, \\
V_2&=-\frac{\hbar^{2}}{8m_2R^{2}}, \;{\text {for}}\; z>0.
\end{align}
This is a well known problem if we ignore the interface effects. This way, the Schr\"{o}dinger equation in both regions can be put as
\begin{align}
&\frac{\partial^{2}\psi_1}{\partial z^{2}}+q_{-}^{2}\psi_1=0,  \;{\text {for}}\; z<0\\
&\frac{\partial^{2}\psi_2}{\partial z^{2}}+q_{+}^{2}\psi_2=0, \;{\text {for}}\; z>0,
\end{align}
where
\begin{eqnarray}
q_-=\sqrt{\frac{2m_1E}{\hbar^{2}}-\frac{ 1}{4R^{2}}-\frac{L_\phi^{2}}{R^{2}}}
\end{eqnarray}
\begin{eqnarray}
q_+=\sqrt{\frac{2m_1E}{\hbar^{2}}-\frac{m_1 }{4m_2R^{2}}-\frac{L_\phi^{2}}{R^{2}}}\;.
\end{eqnarray}
Considering the wave function and its derivative to be continuous everywhere, the reflexion and the transmission coefficient are
\begin{eqnarray}
{\rm R}=\frac{\left(q_--q_+\right)^{2}}{\left(q_-+q_+\right)^{2}}
\end{eqnarray}
and
\begin{eqnarray}
{\rm T}=\frac{4q_-q_+}{\left(q_-+q_+\right)^{2}}\;,
\end{eqnarray}
respectively. We plot them in Fig. \ref{rt}. We chose $m_1=0.4 m_0$ and $m_2=-0.02 m_0$ \cite{negativemass}. As we have said above, the isotropic/anisotropic junctions can be used to manipulate the transport properties in systems modeled by square like quantum potentials in cylindrical geometries \cite{bentcylinder}. Notice that the longitudinal mass components do not change along the entire cylinder: the transmission/reflection phenomena are solely due to the transverse mass component which is present in the geometric potential.  In Ref. \cite{fernandocylinder}, the authors have studied the electronic ballistic transport in more general deformed nanotubes. The anisotropic effects in the geometry induced potential could have significant effects on the electron dynamics in those systems, since they give one more parameter to design the nanotube-based electronic devices. Obviously, other nanotube combinations could be investigated (quantum barrier, quantum well, double quantum well, etc.) in order to search for novel features of the electronic transport in curved structures.
\subsection{Geometry induced potential on a Catenoid}
\begin{figure}[ht]
\begin{center}
   \includegraphics[scale=0.6]{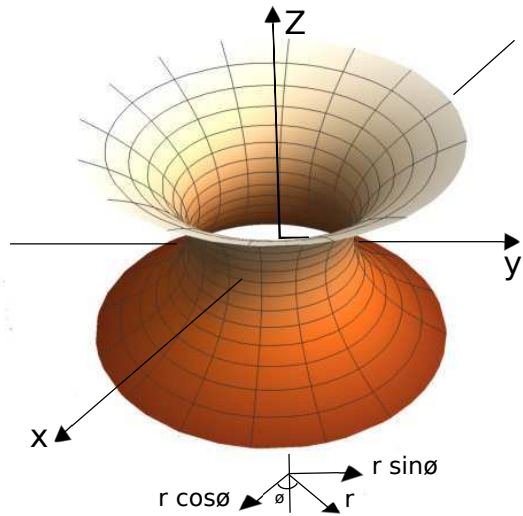}
   \caption{\label{neck} {A catenoid with its axis along $z$ and the throat radius $R$.}}
\end{center}
\end{figure}

\begin{figure}[!htp]
\centering
   \includegraphics[width=0.6\linewidth]{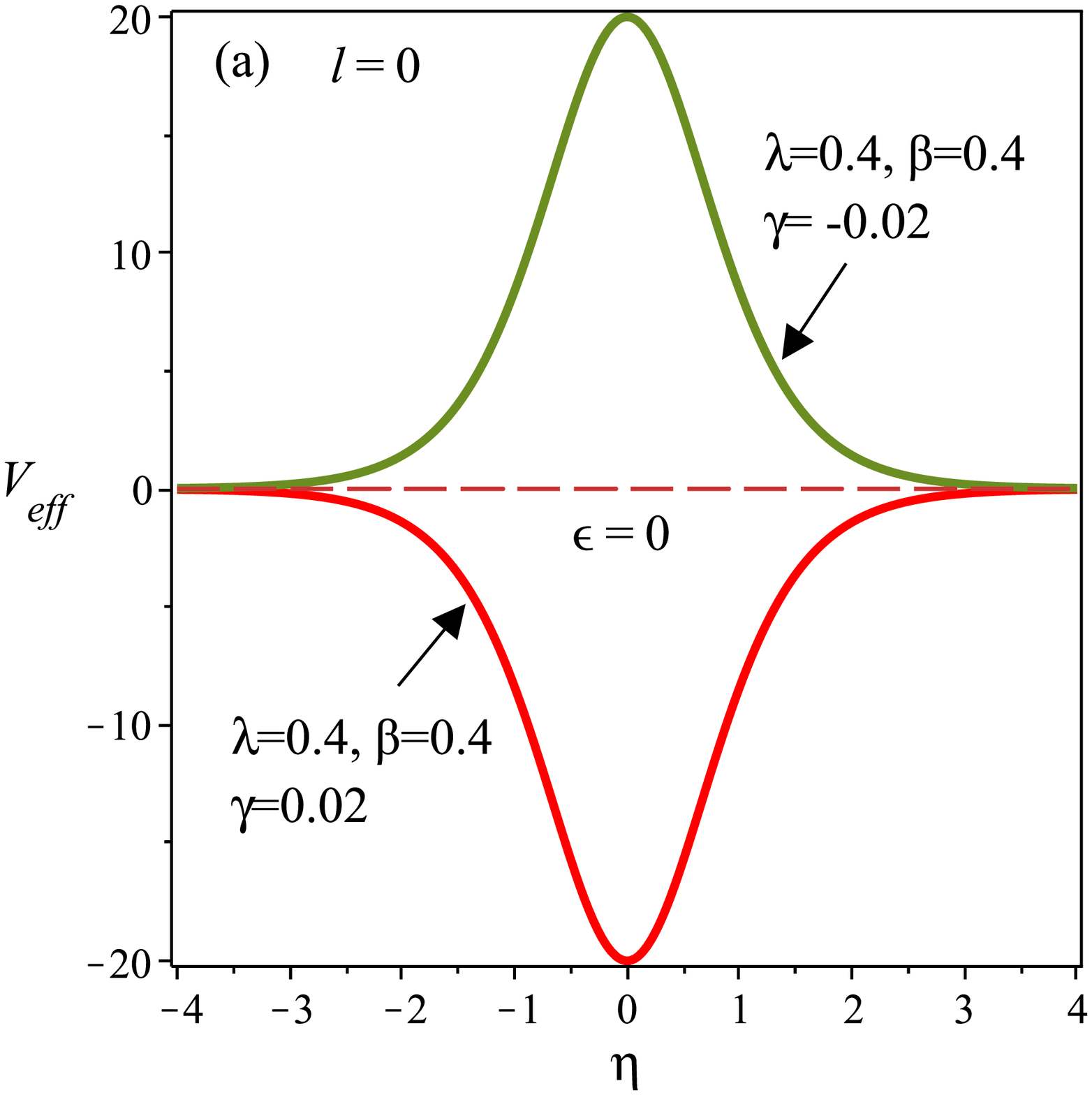}

   \includegraphics[width=0.6\linewidth]{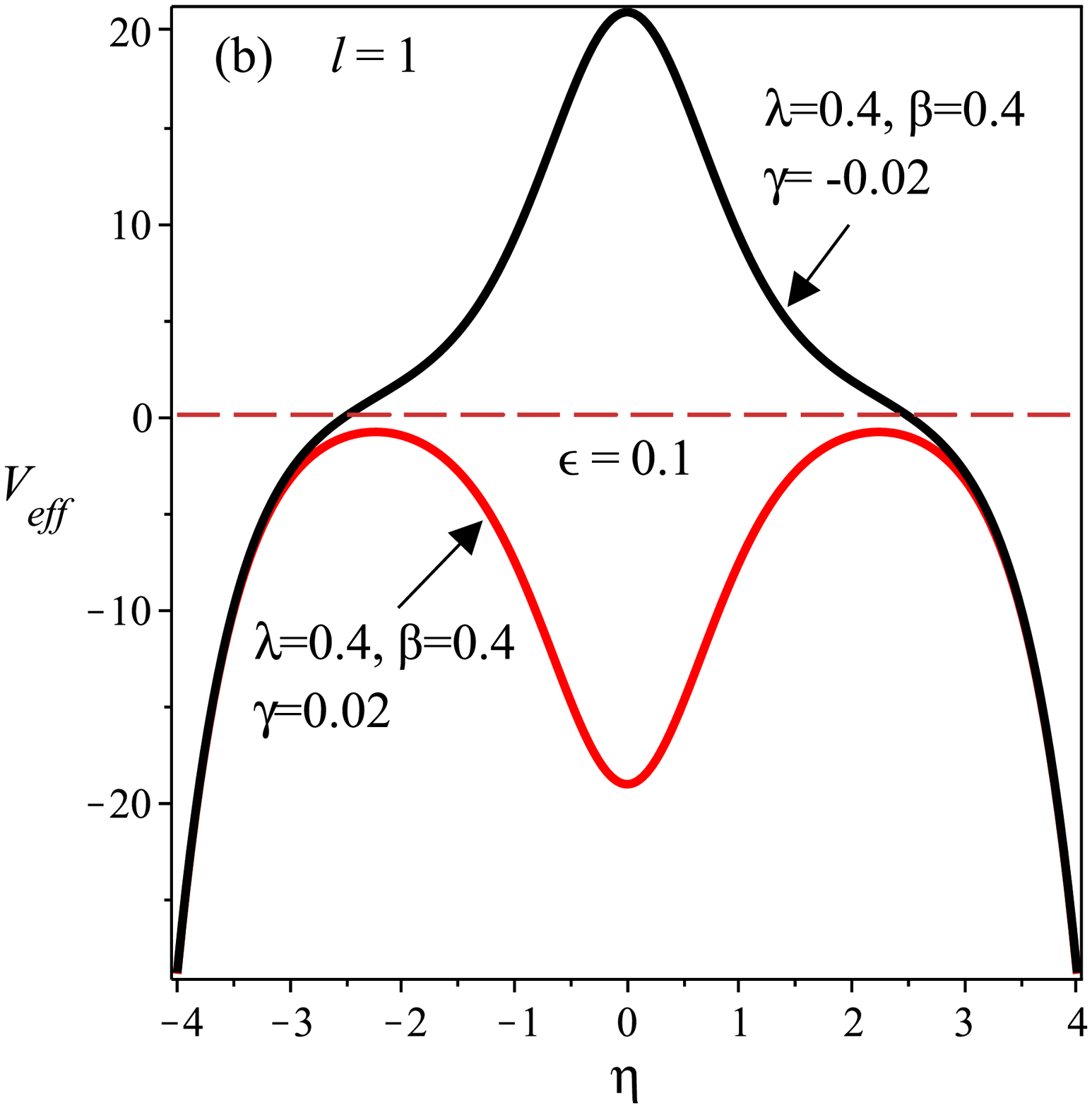}

   \includegraphics[width=0.6\linewidth]{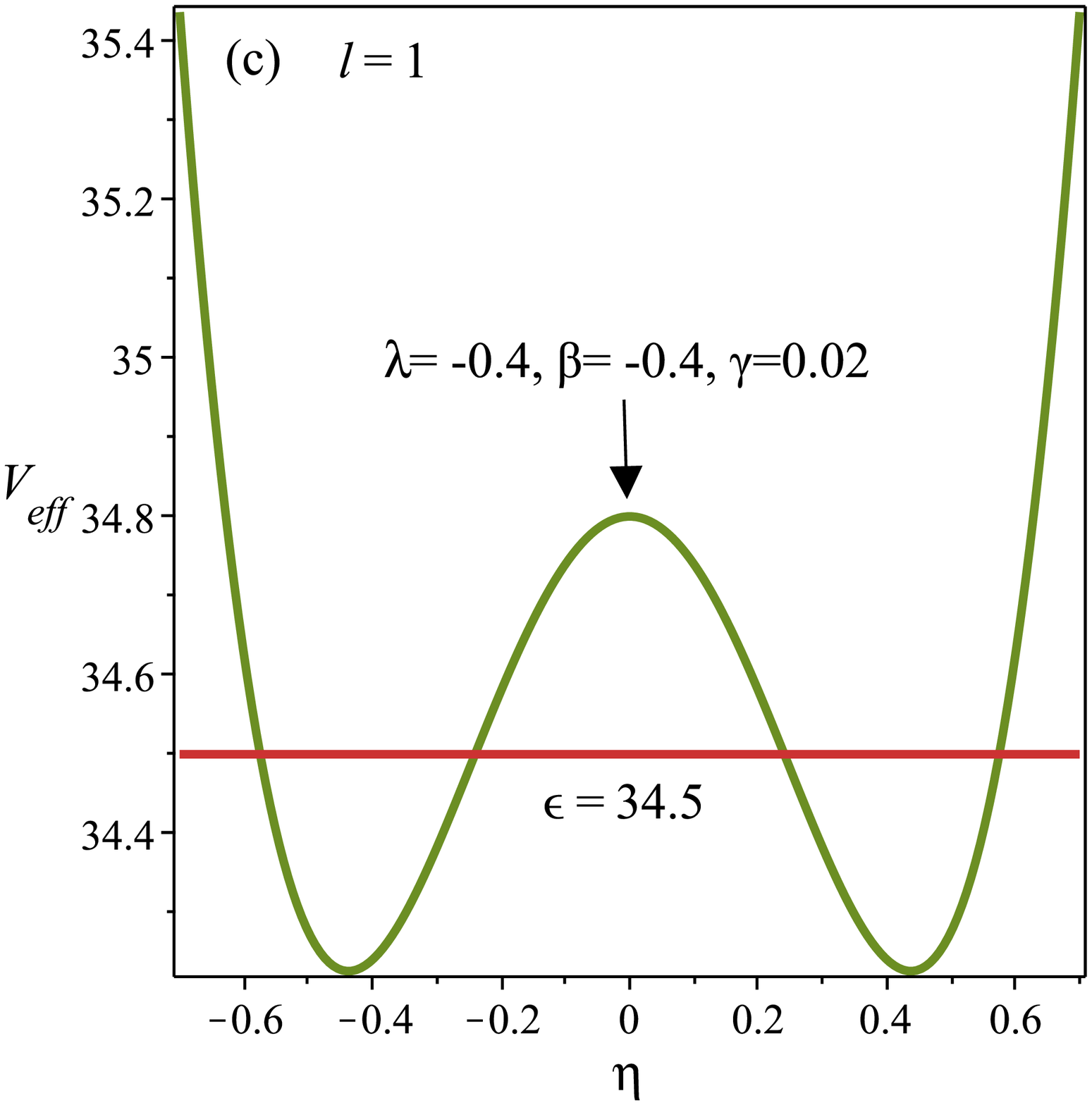}
  \caption{\small The behavior of the curvature induced potential on a catenoid by modifying the ratio between the effective mass components. In 6.({\bf a}), a quantum barrier is yielded when $\epsilon=\ell=0$ for a metamaterial (green), in contrast to the reflectioness potential for an ordinary material (red). In 6.({\bf b}), we see a quantum barrier for a metamaterial and an inverted double well with its maximum lower them the particle energy for an anisotropic ordinary material (we considered $\epsilon=0.1$ and $\ell=1$). In 6.({\bf c}), a double well potential is obtained when one of the surface mass component is negative (we considered $\epsilon=34.5$ and $\ell=1$).}
          \label{cat}
\end{figure}

In f
Fig.\ref{neck}, a catenoid is showed. From it, we define the Cartesian coordinates as  $x=R\cosh(z/R)\cos\phi$, $y=R\cosh(z/R)\sin\phi$ and $z=z$, with $\phi \in [0,2\pi]$.
By considering the $(z,\phi)$ coordinates, the line element will be given by
\begin{equation}
ds^2=\cosh^2\left(\frac{z}{R}\right)dz^2+R^2\cosh^2\left(\frac{z}{R}\right)d\phi^2\;.
\end{equation}
The principal curvatures are
\begin{equation}
\kappa_1=\frac{1}{R}{\rm sech}^2\left(\frac{z}{R}\right), ~~~~\kappa_2=-\frac{1}{R}{\rm sech}^2\left(\frac{z}{R}\right)\;,
\end{equation}
which implies that the mean curvature ${\rm M}=(\kappa_1+\kappa_2)/2=0$ and the Gaussian curvature ${\rm K_G}=\kappa_1\kappa_2=-(1/R^2){\rm sech}^4(z/R)$. This means that the catenoid is a minimal surface \cite{geometry}.
This way, the curvature induced potential for a catenoid will be
\begin{equation}
V(z)=-\frac{\hbar^2}{2m_2}({\rm M^2-K_G})=-\frac{\hbar^2}{2m_2R^2}{\rm sech}^4\left(\frac{z}{R}\right)\;.
\end{equation}
The Schr\"odinger equation is
\begin{eqnarray}
&-&\frac{\hbar^2}{2R\cosh^2 \left(\frac{z}{R} \right)}\left[\frac{R}{m_1}\frac{\partial^2\psi}{\partial z^2}+\frac{1}{m_1^{\prime}R}\frac{\partial^2\psi}{\partial\phi^2}\right] \nonumber\\
&-&\frac{\hbar^2}{2m_2R^2}{\rm sech}^4 \left(\frac{z}{R}\right)\psi=E\psi \;.
\end{eqnarray}
Here, we have considered the general case $m_{1}=\lambda m_{0}$, $m_{1}^{\prime}=\beta m_{0}$ and $m_{2}=\gamma m_{0}$.
Using the cylindrical symmetry along the $z$-axis, setting $\psi=e^{i\ell z}\Phi(z)$, defining dimensionless length $\eta=z/R$ and energy $\epsilon=2m_{0}E
R^2/\hbar^2$, the following effective Schr\"odinger equation is achieved,
\begin{equation}
-\partial^{2}_{\eta\eta}\Phi+V(\eta)\Phi(\eta) = 0\;,  
\end{equation}
where the geometric induced potential reads as
\begin{equation}
V(\eta)=\frac{\lambda}{\beta}\ell^{2}-\frac{\lambda}{\gamma}{\rm sech}^{2}\left(\eta\right)-\lambda\epsilon \cosh^{2}\left(\eta\right)\;,\label{catpotential}\end{equation}
As discussed in \cite{PhysRevA.81.014102} for an isotropic material, this potential for $\ell\neq 0$ is closed related to the corresponding geometric potential for the three dimensional wormhole. For $\epsilon=0$ (ground state) the catenoid enables complete transmission  of a quantum particle across it and the same occurs when we have an anisotropy in the mass with all components being positive (red curve). In the case of a metamaterial (negative transverse mass), this reflectioness potential becomes a quantum barrier (green curve), changing such scenario (see Fig. \ref{cat}({\rm \bf a})). For nonzero and positive $\epsilon$, the potential (\ref{catpotential}) is an inverted double well for an isotropic material and, for an anisotropic ordinary material, a complete transmission of a particle trough the catenoid can be achieved for $\epsilon\neq0$ and $\ell=1$, as showed in Fig. \ref{cat}({\rm \bf b}).  For $\ell\neq0,1$, it can be observed either quantum wells or quantum barriers, depending on the values of the mass components. On the other hand, an interesting case occurs by keeping negative longitudinal mass together with a positive transverse mass: in this case, a double well potential which looks like the one in a double quantum dot \cite{doubledot} is obtained (Fig. \ref{cat}(c)).
\subsection{Quantum Hall effect on a cone}
\subsubsection{Landau levels for a particle on a cone}
Let us consider the coordinates $l$ and $\varphi$ of a particle confined to the surface of a cone  as defined by
\begin{equation}
\left\lbrace
\begin{array}{lll}
$   $x=l\sin\alpha\cos\varphi\\
		y=l\sin\alpha\sin\varphi\\
		z=l\cos\alpha$
$		\end{array}
		\right.\;,\label{cone}
\end{equation}
where $0\leq\varphi\leq 2\pi$ is the usual cylindrical $\varphi$ coordinate and $0<l<+\infty$ (see Fig. \ref{cone}). In the cylindrical coordinates, we have $\rho\equiv l\sin\alpha$.
Consider an applied uniform magnetic field in the $z$-direction, $\vec{B}=(0,0,B_{z})$, as depicted in Fig. \ref{cone}. The vector potential associated to this field is $\vec{A}=\frac{1}{2}\rho B_z \hat{\varphi}$. This way, $\vec{B}=\vec{\nabla}\times \vec{A}=B_{z} \hat{z}$. Then,
\begin{equation}
\vec{A}=\frac{1}{2}B_{z}l\sin\alpha \hat{\varphi}\;.
\end{equation}

\begin{figure}[!htb]
\begin{center}
\includegraphics[height=8cm]{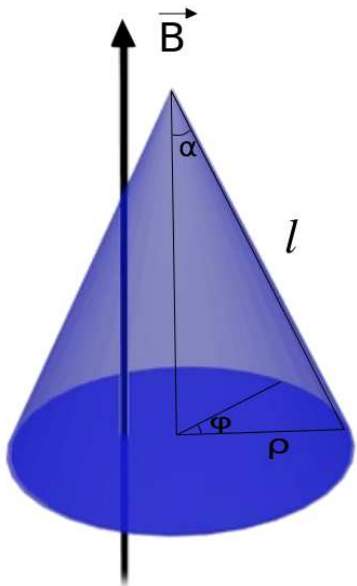}
\caption{Geometrical setting of the problem:   section of height $l\cos\alpha$ of the infinite  straight circular cone of opening angle $2\alpha$ and uniform magnetic field parallel to the cone axis. We consider $\alpha$ between $0$ and $\pi/2$, since, for $\pi/2<\alpha <\pi$ we just have an inverted cone.}
\label{cone}
\end{center}
\end{figure}
By considering the expression (\ref{eq:shrodecop2}) for the Schr\"odinger equation for a charge carrier  confined to the conical surface immersed in an external magnetic field in the $z$-direction, it yields
\begin{align}
-\frac{\hbar^2}{2m_1}\left[\frac{1}{l}\frac{d}{d l}\left(l\frac{d}{d l}\right)-\frac{\frac{1}{4}+\frac{\mu^2}{\sin^2\alpha}}{l^2}\right]\Psi(l)&+\frac{1}{2}m_1w^2l^2\Psi(l)\notag \\&=\Sigma\Psi(l),\label{harm}
\end{align}
where $\mu^2=j^2+\lambda(1-\sin^2\alpha)/4$, with $\lambda=-1$ standing for an ordinary material while $\lambda=-m_1/m_2$ assigns the anisotropic one. We also have $\omega=\left(\omega_c/2 \right)\sin\alpha$, with  $\omega_c=eB_z/m$ (cyclotron frequency) and
\begin{equation}
\Sigma=E+j\hbar\omega_c\;.\label{cond}
\end{equation}
We have considered the wave     function as $\Psi(l,\varphi)=\Psi(l)e^{ij\varphi}$,  with $j=0,\pm1,\pm2,...$ The details of the calculations can be found in \cite{poux2014}. This way, the differential equation (\ref{harm}) describes a two-dimensional quantum oscillator on a conical background. The wave functions are  given by \cite{poux2014}
\begin{equation}
\Psi(l)={\rm C} l^s e^{-\frac{1}{2}m\omega^2 l^2}U\left(\frac{s}{2}+\frac{1}{4}-\frac{E}{2\omega},s+\frac{1}{2},m\omega l^2\right),\label{wf}
\end{equation}
 with
\begin{equation}
s=\frac{1}{2}\left(1\pm\sqrt{1+\frac{4\mu^2}{\sin^2\alpha}}\right)\label{ss}
\end{equation}
and {\rm C} being the normalization constant. $U(a,b,c)$ is the Kummer function \cite{abramo}.

The value of $s$ with a negative sign stands for singular solutions, which have pronounced effects in quantum systems \cite{lasersingular}. It must satisfy the condition $-1<s<1$, which is not achieved in the case of a metamaterial induced repulsive geometric potential on a cone. This is a crucial difference from the ordinary attractive geometric potential. Then, we will take only the case where the s are regular at the conical tip ($l=0$), which is characterized by the following boundary condition,
\begin{equation}
\lim_{l\rightarrow0}l\partial_l \psi(l)=0\;.
\end{equation}
This case means that we do not consider the $\delta$-function potential which comes from the Gaussian curvature of the cone \cite{kowalski}.
In order to have a proper normalization of the wave function, we must have $$\lim_{l\rightarrow\infty}\Psi(l)\rightarrow0\;.$$ In order to get this condition, the series $U\left(\frac{s}{2}+\frac{1}{4}-\frac{E}{2\omega},s+\frac{1}{2},m\omega l^2\right)$ in (\ref{wf}) must be a polynomial of degree n. This is achieved when \cite{abramo}
\begin{equation}
\frac{s}{2}+\frac{1}{4}-\frac{E}{2\omega}=-n\;.
\end{equation}
The corresponding {\it Landau levels } for electrons on a cone are
\begin{equation}
E_{j,n}=\hbar\omega_c\sin\alpha\left(n+\frac{1}{2}+\frac{1}{4}\sqrt{1+\frac{4\mu^2}{\sin^2\alpha}}-\frac{j}{2\sin\alpha}\right).\label{ll}
\end{equation}
\subsubsection{The quantum Hall effect on a cone}
\begin{figure}
\includegraphics[width=0.4 \textwidth]{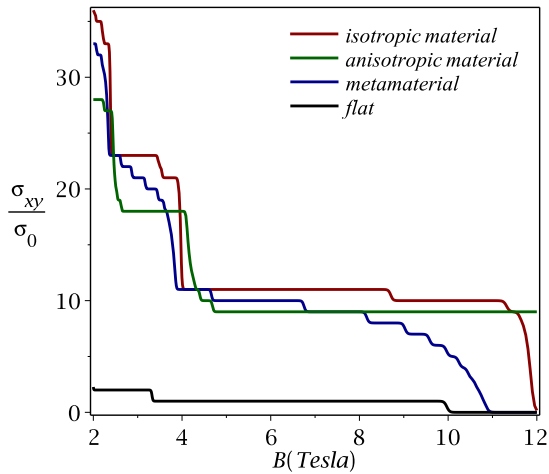}
\caption{Hall conductivity versus the magnetic field on a con  (with a positive curvature). We consider $T=0.3\,K$. Both the plots for an anisotropic material and a metamaterial one are tunned between the curves for a flat isotropic material and a isotropic material in conical shape.} \label{hcone}
\end{figure}

\begin{figure}[!htb]
\includegraphics[width=0.4 \textwidth]{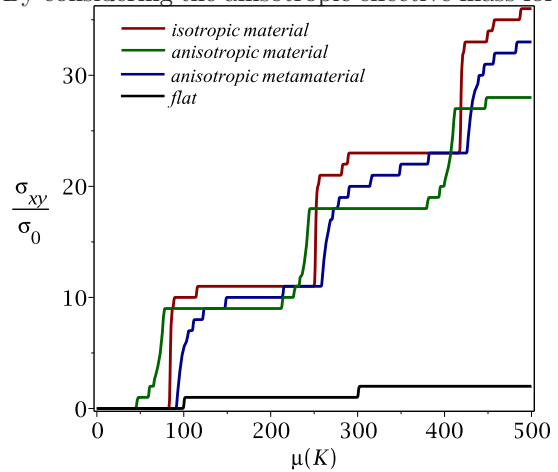}
\caption{Hall conductivity versus the chemical potential on a cone. We consider $B=3T$. Both the plots for an anisotropic material and a metamaterial one are tunned between the curves for a flat isotropic material and a isotropic material in conical shape.} \label{hmu}
\end{figure}

At zero temperature and considering that the Fermi energy $E_F$ is in an energy gap, the Hall conductivity can be written as \cite{Streda}
\begin{equation}
\sigma_H(E_F,0)=\frac{e}{S}\frac{\partial N}{\partial B} \; ,
\end{equation}
where $N$ is the number of states below the Fermi energy and $S$ is the area of the surface. The density of states is given by
\begin{equation}
n(E)=\frac{|eB|}{2\pi \hbar}\sum_{n,l}\delta(E-E_{n,l}) \;.
\end{equation}
$N$ is obtained as
\begin{equation}
N=S\int_{-\infty}^{E_F}n(E)dE=\frac{S|eB|}{h}  n_0,
\end{equation}
where $n_0=(n+1) n_\ell$ is the number of fully occupied LLs below $E_F$, with $n_\ell$ being the number of occupied $\ell$ states. The Hall conductivity is then
\begin{equation}
\sigma_H(E_F,0)=-\frac{e^2}{h}n_o \; .
\end{equation}
For $T\neq0$, we consider the expression for the Hall conductivity obtained in the clean limit (absence of impurities) given by \cite{graphenecondutivity}
{\small\begin{eqnarray}
 \sigma_{H}(\mu,T)=\int_{-\infty}^{\infty}\left(-\frac{\partial f_0}{\partial E}\right)\sigma_{H}(E,0)dE\label{conduc}
\end{eqnarray}}
where $\mu$ is the chemical potential, $T$ is the temperature and $f_0$ is the Fermi-Dirac distribution. We  express the energy scale in  units of temperature. We depict the Hall conductivity versus the magnetic field in Fig. \ref{hcone} and versus the chemical potential in Fig. \ref{hmu}. In the Fig. \ref{hcone}, we plot the Hall conductivity versus the magnetic field intensity. The profile of the Hall curves modify significantly: more plateaus are introduced due to the degeneracy break of the Landau levels and they move to higher magnetic fields for an isotropic material in comparison to the flat sample. The curves for an anisotropic material as well as for a metamaterial one are tunned in between. In Fig. \ref{hmu}, it is depicted the Hall conductivity versus the chemical potential and we observe the profile modifications of it with the plateaus moving to lower values of the chemical potential. Again, the curves for an anisotropic material as well as for a metamaterial one are tunned in between. In addition, it should be noted that the inclusion of the curvature in the system enhances the Hall conductivity in comparison to the case where we have a flat sample.

 The general picture on how the anisotropy on the geometric induced potential is going to affect the Hall system in a conical background can be investigated elsewhere, starting from the studies in Ref. \cite{conePRL}, where the authors have carried out the investigations for ordinary isotropic  materials.
\section{Concluding remarks}\label{sec5}
In this contribution, we have addressed a variant of the problem concerned to the motion of a quantum particle constrained to move on a curved surface. We have employed the thin layer approach introduced by da Costa \cite{dacosta}. By considering the anisotropic effective mass for ballistic electrons, we have showed that the geometric potential changes from the attractive to the repulsive one in some cases, which in turns impact significantly the problems already addressed in the literature for common isotropic materials.
The questions we posed here seems to be simple but we intend to call attention to the works in this field from now on: given the vast possibilities by considering the effective mass theory, new theoretical works should consider variants of the thin layer approach in order to incorporate not only isotropic materials in such investigations, but also the anisotropic ones. The anisotropic mass may have important impact in many other features explored in these curved 2DEG.

{\bf Acknowledgements}: This work was supported by FAPEMIG, CNPq, FAPEMA and CAPES (Brazilian agencies).
\bibliography{sample}

\begin{thebibliography}{44}%
\makeatletter
\providecommand \@ifxundefined [1]{%
 \@ifx{#1\undefined}
}%
\providecommand \@ifnum [1]{%
 \ifnum #1\expandafter \@firstoftwo
 \else \expandafter \@secondoftwo
 \fi
}%
\providecommand \@ifx [1]{%
 \ifx #1\expandafter \@firstoftwo
 \else \expandafter \@secondoftwo
 \fi
}%
\providecommand \natexlab [1]{#1}%
\providecommand \enquote  [1]{``#1''}%
\providecommand \bibnamefont  [1]{#1}%
\providecommand \bibfnamefont [1]{#1}%
\providecommand \citenamefont [1]{#1}%
\providecommand \href@noop [0]{\@secondoftwo}%
\providecommand \href [0]{\begingroup \@sanitize@url \@href}%
\providecommand \@href[1]{\@@startlink{#1}\@@href}%
\providecommand \@@href[1]{\endgroup#1\@@endlink}%
\providecommand \@sanitize@url [0]{\catcode `\\12\catcode `\$12\catcode
  `\&12\catcode `\#12\catcode `\^12\catcode `\_12\catcode `\%12\relax}%
\providecommand \@@startlink[1]{}%
\providecommand \@@endlink[0]{}%
\providecommand \url  [0]{\begingroup\@sanitize@url \@url }%
\providecommand \@url [1]{\endgroup\@href {#1}{\urlprefix }}%
\providecommand \urlprefix  [0]{URL }%
\providecommand \Eprint [0]{\href }%
\providecommand \doibase [0]{http://dx.doi.org/}%
\providecommand \selectlanguage [0]{\@gobble}%
\providecommand \bibinfo  [0]{\@secondoftwo}%
\providecommand \bibfield  [0]{\@secondoftwo}%
\providecommand \translation [1]{[#1]}%
\providecommand \BibitemOpen [0]{}%
\providecommand \bibitemStop [0]{}%
\providecommand \bibitemNoStop [0]{.\EOS\space}%
\providecommand \EOS [0]{\spacefactor3000\relax}%
\providecommand \BibitemShut  [1]{\csname bibitem#1\endcsname}%
\let\auto@bib@innerbib\@empty
\bibitem [{\citenamefont {Cao}\ \emph {et~al.}(2005)\citenamefont {Cao},
  \citenamefont {Laim}, \citenamefont {Ni}, \citenamefont {Nabet},\ and\
  \citenamefont {Spanier}}]{coneexperiment}%
  \BibitemOpen
  \bibfield  {author} {\bibinfo {author} {\bibfnamefont {L.}~\bibnamefont
  {Cao}}, \bibinfo {author} {\bibfnamefont {L.}~\bibnamefont {Laim}}, \bibinfo
  {author} {\bibfnamefont {C.}~\bibnamefont {Ni}}, \bibinfo {author}
  {\bibfnamefont {B.}~\bibnamefont {Nabet}}, \ and\ \bibinfo {author}
  {\bibfnamefont {J.~E.}\ \bibnamefont {Spanier}},\ }\href {\doibase
  10.1021/ja0544814} {\bibfield  {journal} {\bibinfo  {journal} {Journal of the
  American Chemical Society}\ }\textbf {\bibinfo {volume} {127}},\ \bibinfo
  {pages} {13782} (\bibinfo {year} {2005})}\BibitemShut {NoStop}%
\bibitem [{\citenamefont {Prinz}\ \emph {et~al.}(2000)\citenamefont {Prinz},
  \citenamefont {Seleznev}, \citenamefont {Gutakovsky}, \citenamefont
  {Chehovskiy}, \citenamefont {Preobrazhenskii}, \citenamefont {Putyato},\ and\
  \citenamefont {Gavrilova}}]{shapes}%
  \BibitemOpen
  \bibfield  {author} {\bibinfo {author} {\bibfnamefont {V.}~\bibnamefont
  {Prinz}}, \bibinfo {author} {\bibfnamefont {V.}~\bibnamefont {Seleznev}},
  \bibinfo {author} {\bibfnamefont {A.}~\bibnamefont {Gutakovsky}}, \bibinfo
  {author} {\bibfnamefont {A.}~\bibnamefont {Chehovskiy}}, \bibinfo {author}
  {\bibfnamefont {V.}~\bibnamefont {Preobrazhenskii}}, \bibinfo {author}
  {\bibfnamefont {M.}~\bibnamefont {Putyato}}, \ and\ \bibinfo {author}
  {\bibfnamefont {T.}~\bibnamefont {Gavrilova}},\ }\href {\doibase
  http://dx.doi.org/10.1016/S1386-9477(99)00249-0} {\bibfield  {journal}
  {\bibinfo  {journal} {Physica E: Low-dimensional Systems and Nanostructures}\
  }\textbf {\bibinfo {volume} {6}},\ \bibinfo {pages} {828 } (\bibinfo {year}
  {2000})}\BibitemShut {NoStop}%
\bibitem [{\citenamefont {Vorob'ev}\ \emph {et~al.}(2007)\citenamefont
  {Vorob'ev}, \citenamefont {Friedland}, \citenamefont {Kostial}, \citenamefont
  {Hey}, \citenamefont {Jahn}, \citenamefont {Wiebicke}, \citenamefont
  {Yukecheva},\ and\ \citenamefont {Prinz}}]{cylinderexperiment}%
  \BibitemOpen
  \bibfield  {author} {\bibinfo {author} {\bibfnamefont {A.~B.}\ \bibnamefont
  {Vorob'ev}}, \bibinfo {author} {\bibfnamefont {K.-J.}\ \bibnamefont
  {Friedland}}, \bibinfo {author} {\bibfnamefont {H.}~\bibnamefont {Kostial}},
  \bibinfo {author} {\bibfnamefont {R.}~\bibnamefont {Hey}}, \bibinfo {author}
  {\bibfnamefont {U.}~\bibnamefont {Jahn}}, \bibinfo {author} {\bibfnamefont
  {E.}~\bibnamefont {Wiebicke}}, \bibinfo {author} {\bibfnamefont {J.~S.}\
  \bibnamefont {Yukecheva}}, \ and\ \bibinfo {author} {\bibfnamefont {V.~Y.}\
  \bibnamefont {Prinz}},\ }\href {\doibase 10.1103/PhysRevB.75.205309}
  {\bibfield  {journal} {\bibinfo  {journal} {Phys. Rev. B}\ }\textbf {\bibinfo
  {volume} {75}},\ \bibinfo {pages} {205309} (\bibinfo {year}
  {2007})}\BibitemShut {NoStop}%
\bibitem [{\citenamefont {Rosdahl}\ \emph {et~al.}(2015)\citenamefont
  {Rosdahl}, \citenamefont {Manolescu},\ and\ \citenamefont
  {Gudmundsson}}]{introducao3}%
  \BibitemOpen
  \bibfield  {author} {\bibinfo {author} {\bibfnamefont {T.~O.}\ \bibnamefont
  {Rosdahl}}, \bibinfo {author} {\bibfnamefont {A.}~\bibnamefont {Manolescu}},
  \ and\ \bibinfo {author} {\bibfnamefont {V.}~\bibnamefont {Gudmundsson}},\
  }\href {http://dx.doi.org/10.1021/nl503499w} {\bibfield  {journal} {\bibinfo
  {journal} {Nano Letters}\ }\textbf {\bibinfo {volume} {15}},\ \bibinfo
  {pages} {254} (\bibinfo {year} {2015})}\BibitemShut {NoStop}%
\bibitem [{\citenamefont {Can}\ \emph {et~al.}(2014)\citenamefont {Can},
  \citenamefont {Laskin},\ and\ \citenamefont
  {Wiegmann}}]{PhysRevLett.113.046803}%
  \BibitemOpen
  \bibfield  {author} {\bibinfo {author} {\bibfnamefont {T.}~\bibnamefont
  {Can}}, \bibinfo {author} {\bibfnamefont {M.}~\bibnamefont {Laskin}}, \ and\
  \bibinfo {author} {\bibfnamefont {P.}~\bibnamefont {Wiegmann}},\ }\href
  {\doibase 10.1103/PhysRevLett.113.046803} {\bibfield  {journal} {\bibinfo
  {journal} {Phys. Rev. Lett.}\ }\textbf {\bibinfo {volume} {113}},\ \bibinfo
  {pages} {046803} (\bibinfo {year} {2014})}\BibitemShut {NoStop}%
\bibitem [{\citenamefont {da~Costa}(1981)}]{dacosta}%
  \BibitemOpen
  \bibfield  {author} {\bibinfo {author} {\bibfnamefont {R.~C.~T.}\
  \bibnamefont {da~Costa}},\ }\href {\doibase 10.1103/PhysRevA.23.1982}
  {\bibfield  {journal} {\bibinfo  {journal} {Phys. Rev. A}\ }\textbf {\bibinfo
  {volume} {23}},\ \bibinfo {pages} {1982} (\bibinfo {year}
  {1981})}\BibitemShut {NoStop}%
\bibitem [{\citenamefont {Ferrari}\ and\ \citenamefont
  {Cuoghi}(2008)}]{PhysRevLett.100.230403}%
  \BibitemOpen
  \bibfield  {author} {\bibinfo {author} {\bibfnamefont {G.}~\bibnamefont
  {Ferrari}}\ and\ \bibinfo {author} {\bibfnamefont {G.}~\bibnamefont
  {Cuoghi}},\ }\href {\doibase 10.1103/PhysRevLett.100.230403} {\bibfield
  {journal} {\bibinfo  {journal} {Phys. Rev. Lett.}\ }\textbf {\bibinfo
  {volume} {100}},\ \bibinfo {pages} {230403} (\bibinfo {year}
  {2008})}\BibitemShut {NoStop}%
\bibitem [{\citenamefont {Atanasov}\ and\ \citenamefont
  {Saxena}(2015)}]{PhysRevB.92.035440}%
  \BibitemOpen
  \bibfield  {author} {\bibinfo {author} {\bibfnamefont {V.}~\bibnamefont
  {Atanasov}}\ and\ \bibinfo {author} {\bibfnamefont {A.}~\bibnamefont
  {Saxena}},\ }\href {\doibase 10.1103/PhysRevB.92.035440} {\bibfield
  {journal} {\bibinfo  {journal} {Phys. Rev. B}\ }\textbf {\bibinfo {volume}
  {92}},\ \bibinfo {pages} {035440} (\bibinfo {year} {2015})}\BibitemShut
  {NoStop}%
\bibitem [{\citenamefont {Can}\ \emph {et~al.}(2016)\citenamefont {Can},
  \citenamefont {Chiu}, \citenamefont {Laskin},\ and\ \citenamefont
  {Wiegmann}}]{conePRL}%
  \BibitemOpen
  \bibfield  {author} {\bibinfo {author} {\bibfnamefont {T.}~\bibnamefont
  {Can}}, \bibinfo {author} {\bibfnamefont {Y.~H.}\ \bibnamefont {Chiu}},
  \bibinfo {author} {\bibfnamefont {M.}~\bibnamefont {Laskin}}, \ and\ \bibinfo
  {author} {\bibfnamefont {P.}~\bibnamefont {Wiegmann}},\ }\href {\doibase
  10.1103/PhysRevLett.117.266803} {\bibfield  {journal} {\bibinfo  {journal}
  {Phys. Rev. Lett.}\ }\textbf {\bibinfo {volume} {117}},\ \bibinfo {pages}
  {266803} (\bibinfo {year} {2016})}\BibitemShut {NoStop}%
\bibitem [{\citenamefont {Wang}\ \emph {et~al.}(2014)\citenamefont {Wang},
  \citenamefont {Du}, \citenamefont {Xu}, \citenamefont {Liu},\ and\
  \citenamefont {Zong}}]{paulicurved}%
  \BibitemOpen
  \bibfield  {author} {\bibinfo {author} {\bibfnamefont {Y.-L.}\ \bibnamefont
  {Wang}}, \bibinfo {author} {\bibfnamefont {L.}~\bibnamefont {Du}}, \bibinfo
  {author} {\bibfnamefont {C.-T.}\ \bibnamefont {Xu}}, \bibinfo {author}
  {\bibfnamefont {X.-J.}\ \bibnamefont {Liu}}, \ and\ \bibinfo {author}
  {\bibfnamefont {H.-S.}\ \bibnamefont {Zong}},\ }\href {\doibase
  10.1103/PhysRevA.90.042117} {\bibfield  {journal} {\bibinfo  {journal} {Phys.
  Rev. A}\ }\textbf {\bibinfo {volume} {90}},\ \bibinfo {pages} {042117}
  (\bibinfo {year} {2014})}\BibitemShut {NoStop}%
\bibitem [{\citenamefont {Wang}\ \emph {et~al.}(2017)\citenamefont {Wang},
  \citenamefont {Jiang},\ and\ \citenamefont {Zong}}]{truncado}%
  \BibitemOpen
  \bibfield  {author} {\bibinfo {author} {\bibfnamefont {Y.-L.}\ \bibnamefont
  {Wang}}, \bibinfo {author} {\bibfnamefont {H.}~\bibnamefont {Jiang}}, \ and\
  \bibinfo {author} {\bibfnamefont {H.-S.}\ \bibnamefont {Zong}},\ }\href
  {\doibase 10.1103/PhysRevA.96.022116} {\bibfield  {journal} {\bibinfo
  {journal} {Phys. Rev. A}\ }\textbf {\bibinfo {volume} {96}},\ \bibinfo
  {pages} {022116} (\bibinfo {year} {2017})}\BibitemShut {NoStop}%
\bibitem [{\citenamefont {Wang}\ and\ \citenamefont {Zong}(2016)}]{Wang201668}%
  \BibitemOpen
  \bibfield  {author} {\bibinfo {author} {\bibfnamefont {Y.-L.}\ \bibnamefont
  {Wang}}\ and\ \bibinfo {author} {\bibfnamefont {H.-S.}\ \bibnamefont
  {Zong}},\ }\href {\doibase 10.1016/j.aop.2015.10.019} {\bibfield  {journal}
  {\bibinfo  {journal} {Annals of Physics}\ }\textbf {\bibinfo {volume}
  {364}},\ \bibinfo {pages} {68} (\bibinfo {year} {2016})}\BibitemShut
  {NoStop}%
\bibitem [{\citenamefont {Xun}\ and\ \citenamefont {Liu}(2014)}]{Xun2014132}%
  \BibitemOpen
  \bibfield  {author} {\bibinfo {author} {\bibfnamefont {D.}~\bibnamefont
  {Xun}}\ and\ \bibinfo {author} {\bibfnamefont {Q.}~\bibnamefont {Liu}},\
  }\href {\doibase http://dx.doi.org/10.1016/j.aop.2013.11.017} {\bibfield
  {journal} {\bibinfo  {journal} {Annals of Physics}\ }\textbf {\bibinfo
  {volume} {341}},\ \bibinfo {pages} {132 } (\bibinfo {year}
  {2014})}\BibitemShut {NoStop}%
\bibitem [{\citenamefont {Spittel}\ \emph {et~al.}(2015)\citenamefont
  {Spittel}, \citenamefont {Uebel}, \citenamefont {Bartelt},\ and\
  \citenamefont {Schmidt}}]{Spittel:15}%
  \BibitemOpen
  \bibfield  {author} {\bibinfo {author} {\bibfnamefont {R.}~\bibnamefont
  {Spittel}}, \bibinfo {author} {\bibfnamefont {P.}~\bibnamefont {Uebel}},
  \bibinfo {author} {\bibfnamefont {H.}~\bibnamefont {Bartelt}}, \ and\
  \bibinfo {author} {\bibfnamefont {M.~A.}\ \bibnamefont {Schmidt}},\ }\href
  {\doibase 10.1364/OE.23.012174} {\bibfield  {journal} {\bibinfo  {journal}
  {Opt. Express}\ }\textbf {\bibinfo {volume} {23}},\ \bibinfo {pages} {12174}
  (\bibinfo {year} {2015})}\BibitemShut {NoStop}%
\bibitem [{\citenamefont {Jahangiri}\ and\ \citenamefont
  {Panahi}(2016)}]{Jahangiri2016407}%
  \BibitemOpen
  \bibfield  {author} {\bibinfo {author} {\bibfnamefont {L.}~\bibnamefont
  {Jahangiri}}\ and\ \bibinfo {author} {\bibfnamefont {H.}~\bibnamefont
  {Panahi}},\ }\href {\doibase http://dx.doi.org/10.1016/j.aop.2016.10.011}
  {\bibfield  {journal} {\bibinfo  {journal} {Annals of Physics}\ }\textbf
  {\bibinfo {volume} {375}},\ \bibinfo {pages} {407 } (\bibinfo {year}
  {2016})}\BibitemShut {NoStop}%
\bibitem [{\citenamefont {Panahi}\ and\ \citenamefont
  {Jahangiri}(2016)}]{Panahi201657}%
  \BibitemOpen
  \bibfield  {author} {\bibinfo {author} {\bibfnamefont {H.}~\bibnamefont
  {Panahi}}\ and\ \bibinfo {author} {\bibfnamefont {L.}~\bibnamefont
  {Jahangiri}},\ }\href {\doibase 10.1016/j.aop.2016.04.013} {\bibfield
  {journal} {\bibinfo  {journal} {Annals of Physics}\ }\textbf {\bibinfo
  {volume} {372}},\ \bibinfo {pages} {57} (\bibinfo {year} {2016})},\ \bibinfo
  {note} {cited By 1}\BibitemShut {NoStop}%
\bibitem [{\citenamefont {Wang}\ \emph {et~al.}(2016)\citenamefont {Wang},
  \citenamefont {Liang}, \citenamefont {Jiang}, \citenamefont {Lu},\ and\
  \citenamefont {Zong}}]{Wang2016}%
  \BibitemOpen
  \bibfield  {author} {\bibinfo {author} {\bibfnamefont {Y.-L.}\ \bibnamefont
  {Wang}}, \bibinfo {author} {\bibfnamefont {G.-H.}\ \bibnamefont {Liang}},
  \bibinfo {author} {\bibfnamefont {H.}~\bibnamefont {Jiang}}, \bibinfo
  {author} {\bibfnamefont {W.-T.}\ \bibnamefont {Lu}}, \ and\ \bibinfo {author}
  {\bibfnamefont {H.-S.}\ \bibnamefont {Zong}},\ }\href {\doibase
  10.1088/0022-3727/49/29/295103} {\bibfield  {journal} {\bibinfo  {journal}
  {Journal of Physics D: Applied Physics}\ }\textbf {\bibinfo {volume} {49}}
  (\bibinfo {year} {2016}),\ 10.1088/0022-3727/49/29/295103},\ \bibinfo {note}
  {cited By 0}\BibitemShut {NoStop}%
\bibitem [{\citenamefont {Cruz}\ \emph {et~al.}(2017)\citenamefont {Cruz},
  \citenamefont {Bernardo},\ and\ \citenamefont {Esguerra}}]{Cruz2017}%
  \BibitemOpen
  \bibfield  {author} {\bibinfo {author} {\bibfnamefont {P.~C.~S.}\
  \bibnamefont {Cruz}}, \bibinfo {author} {\bibfnamefont {R.~C.~S.}\
  \bibnamefont {Bernardo}}, \ and\ \bibinfo {author} {\bibfnamefont {J.~P.~H.}\
  \bibnamefont {Esguerra}},\ }\href {\doibase
  http://dx.doi.org/10.1016/j.aop.2017.02.004} {\bibfield  {journal} {\bibinfo
  {journal} {Annals of Physics}\ ,\ } (\bibinfo {year} {2017})}\BibitemShut
  {NoStop}%
\bibitem [{\citenamefont {Liu}\ \emph {et~al.}(2017)\citenamefont {Liu},
  \citenamefont {Zhang}, \citenamefont {Lian}, \citenamefont {Hu},\ and\
  \citenamefont {Li}}]{centripetal}%
  \BibitemOpen
  \bibfield  {author} {\bibinfo {author} {\bibfnamefont {Q.}~\bibnamefont
  {Liu}}, \bibinfo {author} {\bibfnamefont {J.}~\bibnamefont {Zhang}}, \bibinfo
  {author} {\bibfnamefont {D.}~\bibnamefont {Lian}}, \bibinfo {author}
  {\bibfnamefont {L.}~\bibnamefont {Hu}}, \ and\ \bibinfo {author}
  {\bibfnamefont {Z.}~\bibnamefont {Li}},\ }\href {\doibase
  https://doi.org/10.1016/j.physe.2016.11.029} {\bibfield  {journal} {\bibinfo
  {journal} {Physica E: Low-dimensional Systems and Nanostructures}\ }\textbf
  {\bibinfo {volume} {87}},\ \bibinfo {pages} {123 } (\bibinfo {year}
  {2017})}\BibitemShut {NoStop}%
\bibitem [{\citenamefont {Kittel}(2005)}]{kittel2005introduction}%
  \BibitemOpen
  \bibfield  {author} {\bibinfo {author} {\bibfnamefont {C.}~\bibnamefont
  {Kittel}},\ }\href@noop {} {\emph {\bibinfo {title} {Introduction to solid
  state physics}}},\ \bibinfo {edition} {8th}\ ed.\ (\bibinfo  {publisher}
  {Wiley},\ \bibinfo {year} {2005})\BibitemShut {NoStop}%
\bibitem [{\citenamefont {Silveirinha}\ and\ \citenamefont
  {Engheta}(2012{\natexlab{a}})}]{tailoring}%
  \BibitemOpen
  \bibfield  {author} {\bibinfo {author} {\bibfnamefont {M.~G.}\ \bibnamefont
  {Silveirinha}}\ and\ \bibinfo {author} {\bibfnamefont {N.}~\bibnamefont
  {Engheta}},\ }\href {\doibase 10.1103/PhysRevB.86.161104} {\bibfield
  {journal} {\bibinfo  {journal} {Phys. Rev. B}\ }\textbf {\bibinfo {volume}
  {86}},\ \bibinfo {pages} {161104} (\bibinfo {year}
  {2012}{\natexlab{a}})}\BibitemShut {NoStop}%
\bibitem [{\citenamefont {Silveirinha}\ and\ \citenamefont
  {Engheta}(2012{\natexlab{b}})}]{silverinha}%
  \BibitemOpen
  \bibfield  {author} {\bibinfo {author} {\bibfnamefont {M.~G.}\ \bibnamefont
  {Silveirinha}}\ and\ \bibinfo {author} {\bibfnamefont {N.}~\bibnamefont
  {Engheta}},\ }\href {\doibase 10.1103/PhysRevB.86.245302} {\bibfield
  {journal} {\bibinfo  {journal} {Phys. Rev. B}\ }\textbf {\bibinfo {volume}
  {86}},\ \bibinfo {pages} {245302} (\bibinfo {year}
  {2012}{\natexlab{b}})}\BibitemShut {NoStop}%
\bibitem [{\citenamefont {Yukawa}\ \emph {et~al.}(2015)\citenamefont {Yukawa},
  \citenamefont {Ozawa}, \citenamefont {Yamamoto}, \citenamefont {Liu},\ and\
  \citenamefont {Matsuda}}]{anisotropicmass}%
  \BibitemOpen
  \bibfield  {author} {\bibinfo {author} {\bibfnamefont {R.}~\bibnamefont
  {Yukawa}}, \bibinfo {author} {\bibfnamefont {K.}~\bibnamefont {Ozawa}},
  \bibinfo {author} {\bibfnamefont {S.}~\bibnamefont {Yamamoto}}, \bibinfo
  {author} {\bibfnamefont {R.-Y.}\ \bibnamefont {Liu}}, \ and\ \bibinfo
  {author} {\bibfnamefont {I.}~\bibnamefont {Matsuda}},\ }\href {\doibase
  https://doi.org/10.1016/j.susc.2015.07.007} {\bibfield  {journal} {\bibinfo
  {journal} {Surface Science}\ }\textbf {\bibinfo {volume} {641}},\ \bibinfo
  {pages} {224 } (\bibinfo {year} {2015})}\BibitemShut {NoStop}%
\bibitem [{\citenamefont {Dragoman}\ and\ \citenamefont
  {Dragoman}(2007)}]{dragoman2007metamaterials}%
  \BibitemOpen
  \bibfield  {author} {\bibinfo {author} {\bibfnamefont {D.}~\bibnamefont
  {Dragoman}}\ and\ \bibinfo {author} {\bibfnamefont {M.}~\bibnamefont
  {Dragoman}},\ }\href@noop {} {\bibfield  {journal} {\bibinfo  {journal}
  {Journal of Applied Physics}\ }\textbf {\bibinfo {volume} {101}},\ \bibinfo
  {pages} {104316} (\bibinfo {year} {2007})}\BibitemShut {NoStop}%
\bibitem [{\citenamefont {Walia}\ \emph {et~al.}(2015)\citenamefont {Walia},
  \citenamefont {Shah}, \citenamefont {Gutruf}, \citenamefont {Nili},
  \citenamefont {Chowdhury}, \citenamefont {Withayachumnankul}, \citenamefont
  {Bhaskaran},\ and\ \citenamefont {Sriram}}]{metareview}%
  \BibitemOpen
  \bibfield  {author} {\bibinfo {author} {\bibfnamefont {S.}~\bibnamefont
  {Walia}}, \bibinfo {author} {\bibfnamefont {C.~M.}\ \bibnamefont {Shah}},
  \bibinfo {author} {\bibfnamefont {P.}~\bibnamefont {Gutruf}}, \bibinfo
  {author} {\bibfnamefont {H.}~\bibnamefont {Nili}}, \bibinfo {author}
  {\bibfnamefont {D.~R.}\ \bibnamefont {Chowdhury}}, \bibinfo {author}
  {\bibfnamefont {W.}~\bibnamefont {Withayachumnankul}}, \bibinfo {author}
  {\bibfnamefont {M.}~\bibnamefont {Bhaskaran}}, \ and\ \bibinfo {author}
  {\bibfnamefont {S.}~\bibnamefont {Sriram}},\ }\href {\doibase
  10.1063/1.4913751} {\bibfield  {journal} {\bibinfo  {journal} {Applied
  Physics Reviews}\ }\textbf {\bibinfo {volume} {2}},\ \bibinfo {pages}
  {011303} (\bibinfo {year} {2015})}\BibitemShut {NoStop}%
\bibitem [{\citenamefont {Mol}\ and\ \citenamefont
  {Aanandan}(2017)}]{JPCOmeta}%
  \BibitemOpen
  \bibfield  {author} {\bibinfo {author} {\bibfnamefont {V.~A.~L.}\
  \bibnamefont {Mol}}\ and\ \bibinfo {author} {\bibfnamefont {C.~K.}\
  \bibnamefont {Aanandan}},\ }\href
  {http://stacks.iop.org/2399-6528/1/i=1/a=015003} {\bibfield  {journal}
  {\bibinfo  {journal} {Journal of Physics Communications}\ }\textbf {\bibinfo
  {volume} {1}},\ \bibinfo {pages} {015003} (\bibinfo {year}
  {2017})}\BibitemShut {NoStop}%
\bibitem [{\citenamefont {Figueiredo}\ \emph {et~al.}(2016)\citenamefont
  {Figueiredo}, \citenamefont {Gomes}, \citenamefont {Fumeron}, \citenamefont
  {Berche},\ and\ \citenamefont {Moraes}}]{PhysRevD.94.044039}%
  \BibitemOpen
  \bibfield  {author} {\bibinfo {author} {\bibfnamefont {D.}~\bibnamefont
  {Figueiredo}}, \bibinfo {author} {\bibfnamefont {F.~A.}\ \bibnamefont
  {Gomes}}, \bibinfo {author} {\bibfnamefont {S.}~\bibnamefont {Fumeron}},
  \bibinfo {author} {\bibfnamefont {B.}~\bibnamefont {Berche}}, \ and\ \bibinfo
  {author} {\bibfnamefont {F.}~\bibnamefont {Moraes}},\ }\href {\doibase
  10.1103/PhysRevD.94.044039} {\bibfield  {journal} {\bibinfo  {journal} {Phys.
  Rev. D}\ }\textbf {\bibinfo {volume} {94}},\ \bibinfo {pages} {044039}
  (\bibinfo {year} {2016})}\BibitemShut {NoStop}%
\bibitem [{\citenamefont {Shekhar}\ \emph {et~al.}(2014)\citenamefont
  {Shekhar}, \citenamefont {Atkinson},\ and\ \citenamefont
  {Jacob}}]{shekhar2014hyperbolic}%
  \BibitemOpen
  \bibfield  {author} {\bibinfo {author} {\bibfnamefont {P.}~\bibnamefont
  {Shekhar}}, \bibinfo {author} {\bibfnamefont {J.}~\bibnamefont {Atkinson}}, \
  and\ \bibinfo {author} {\bibfnamefont {Z.}~\bibnamefont {Jacob}},\
  }\href@noop {} {\bibfield  {journal} {\bibinfo  {journal} {Nano Convergence}\
  }\textbf {\bibinfo {volume} {1}},\ \bibinfo {pages} {1} (\bibinfo {year}
  {2014})}\BibitemShut {NoStop}%
\bibitem [{\citenamefont {Weisstein}(2008)}]{Wolfram}%
  \BibitemOpen
  \bibfield  {author} {\bibinfo {author} {\bibfnamefont {E.}~\bibnamefont
  {Weisstein}},\ }\href@noop {} {\bibfield  {journal} {\bibinfo  {journal}
  {{\it Weingarten Equations}}\ ,\ \bibinfo {pages}
  {http://mathworld.wolfram.com/WeingartenEquations.html}} (\bibinfo {year}
  {2008})}\BibitemShut {NoStop}%
\bibitem [{\citenamefont {Dyson}\ and\ \citenamefont
  {Ridley}(2005)}]{negativemass}%
  \BibitemOpen
  \bibfield  {author} {\bibinfo {author} {\bibfnamefont {A.}~\bibnamefont
  {Dyson}}\ and\ \bibinfo {author} {\bibfnamefont {B.~K.}\ \bibnamefont
  {Ridley}},\ }\href {\doibase 10.1103/PhysRevB.72.193301} {\bibfield
  {journal} {\bibinfo  {journal} {Phys. Rev. B}\ }\textbf {\bibinfo {volume}
  {72}},\ \bibinfo {pages} {193301} (\bibinfo {year} {2005})}\BibitemShut
  {NoStop}%
\bibitem [{\citenamefont {Gray}(1993)}]{geometry}%
  \BibitemOpen
  \bibfield  {author} {\bibinfo {author} {\bibfnamefont {A.}~\bibnamefont
  {Gray}},\ }\href@noop {} {\emph {\bibinfo {title} {Modern Differential
  Geometry of Curves and Surfaces}}}\ (\bibinfo  {publisher} {Boca Raton:CRC
  Press},\ \bibinfo {year} {1993})\BibitemShut {NoStop}%
\bibitem [{\citenamefont {Atanasov}\ \emph {et~al.}(2009)\citenamefont
  {Atanasov}, \citenamefont {Dandoloff},\ and\ \citenamefont
  {Saxena}}]{atanasov}%
  \BibitemOpen
  \bibfield  {author} {\bibinfo {author} {\bibfnamefont {V.}~\bibnamefont
  {Atanasov}}, \bibinfo {author} {\bibfnamefont {R.}~\bibnamefont {Dandoloff}},
  \ and\ \bibinfo {author} {\bibfnamefont {A.}~\bibnamefont {Saxena}},\ }\href
  {\doibase 10.1103/PhysRevB.79.033404} {\bibfield  {journal} {\bibinfo
  {journal} {Phys. Rev. B}\ }\textbf {\bibinfo {volume} {79}},\ \bibinfo
  {pages} {033404} (\bibinfo {year} {2009})}\BibitemShut {NoStop}%
\bibitem [{\citenamefont {Rezaei}\ \emph {et~al.}(2011)\citenamefont {Rezaei},
  \citenamefont {Vaseghi}, \citenamefont {Khordad},\ and\ \citenamefont
  {Kenary}}]{rectification}%
  \BibitemOpen
  \bibfield  {author} {\bibinfo {author} {\bibfnamefont {G.}~\bibnamefont
  {Rezaei}}, \bibinfo {author} {\bibfnamefont {B.}~\bibnamefont {Vaseghi}},
  \bibinfo {author} {\bibfnamefont {R.}~\bibnamefont {Khordad}}, \ and\
  \bibinfo {author} {\bibfnamefont {H.~A.}\ \bibnamefont {Kenary}},\ }\href
  {\doibase https://doi.org/10.1016/j.physe.2011.06.026} {\bibfield  {journal}
  {\bibinfo  {journal} {Physica E: Low-dimensional Systems and Nanostructures}\
  }\textbf {\bibinfo {volume} {43}},\ \bibinfo {pages} {1853 } (\bibinfo {year}
  {2011})}\BibitemShut {NoStop}%
\bibitem [{\citenamefont {Bejan}\ \emph {et~al.}(2018)\citenamefont {Bejan},
  \citenamefont {Stan},\ and\ \citenamefont {Niculescu}}]{secthird}%
  \BibitemOpen
  \bibfield  {author} {\bibinfo {author} {\bibfnamefont {D.}~\bibnamefont
  {Bejan}}, \bibinfo {author} {\bibfnamefont {C.}~\bibnamefont {Stan}}, \ and\
  \bibinfo {author} {\bibfnamefont {E.~C.}\ \bibnamefont {Niculescu}},\ }\href
  {\doibase https://doi.org/10.1016/j.optmat.2018.02.008} {\bibfield  {journal}
  {\bibinfo  {journal} {Optical Materials}\ }\textbf {\bibinfo {volume} {78}},\
  \bibinfo {pages} {207 } (\bibinfo {year} {2018})}\BibitemShut {NoStop}%
\bibitem [{\citenamefont {Marchi}\ \emph {et~al.}(2005)\citenamefont {Marchi},
  \citenamefont {Reggiani}, \citenamefont {Rudan},\ and\ \citenamefont
  {Bertoni}}]{bentcylinder}%
  \BibitemOpen
  \bibfield  {author} {\bibinfo {author} {\bibfnamefont {A.}~\bibnamefont
  {Marchi}}, \bibinfo {author} {\bibfnamefont {S.}~\bibnamefont {Reggiani}},
  \bibinfo {author} {\bibfnamefont {M.}~\bibnamefont {Rudan}}, \ and\ \bibinfo
  {author} {\bibfnamefont {A.}~\bibnamefont {Bertoni}},\ }\href {\doibase
  10.1103/PhysRevB.72.035403} {\bibfield  {journal} {\bibinfo  {journal} {Phys.
  Rev. B}\ }\textbf {\bibinfo {volume} {72}},\ \bibinfo {pages} {035403}
  (\bibinfo {year} {2005})}\BibitemShut {NoStop}%
\bibitem [{\citenamefont {Santos}\ \emph {et~al.}(2016)\citenamefont {Santos},
  \citenamefont {Fumeron}, \citenamefont {Berche},\ and\ \citenamefont
  {Moraes}}]{fernandocylinder}%
  \BibitemOpen
  \bibfield  {author} {\bibinfo {author} {\bibfnamefont {F.}~\bibnamefont
  {Santos}}, \bibinfo {author} {\bibfnamefont {S.}~\bibnamefont {Fumeron}},
  \bibinfo {author} {\bibfnamefont {B.}~\bibnamefont {Berche}}, \ and\ \bibinfo
  {author} {\bibfnamefont {F.}~\bibnamefont {Moraes}},\ }\href
  {http://stacks.iop.org/0957-4484/27/i=13/a=135302} {\bibfield  {journal}
  {\bibinfo  {journal} {Nanotechnology}\ }\textbf {\bibinfo {volume} {27}},\
  \bibinfo {pages} {135302} (\bibinfo {year} {2016})}\BibitemShut {NoStop}%
\bibitem [{\citenamefont {Dandoloff}\ \emph {et~al.}(2010)\citenamefont
  {Dandoloff}, \citenamefont {Saxena},\ and\ \citenamefont
  {Jensen}}]{PhysRevA.81.014102}%
  \BibitemOpen
  \bibfield  {author} {\bibinfo {author} {\bibfnamefont {R.}~\bibnamefont
  {Dandoloff}}, \bibinfo {author} {\bibfnamefont {A.}~\bibnamefont {Saxena}}, \
  and\ \bibinfo {author} {\bibfnamefont {B.}~\bibnamefont {Jensen}},\ }\href
  {\doibase 10.1103/PhysRevA.81.014102} {\bibfield  {journal} {\bibinfo
  {journal} {Phys. Rev. A}\ }\textbf {\bibinfo {volume} {81}},\ \bibinfo
  {pages} {014102} (\bibinfo {year} {2010})}\BibitemShut {NoStop}%
\bibitem [{\citenamefont {Fanchini}\ \emph {et~al.}(2010)\citenamefont
  {Fanchini}, \citenamefont {Castelano},\ and\ \citenamefont
  {Caldeira}}]{doubledot}%
  \BibitemOpen
  \bibfield  {author} {\bibinfo {author} {\bibfnamefont {F.~F.}\ \bibnamefont
  {Fanchini}}, \bibinfo {author} {\bibfnamefont {L.~K.}\ \bibnamefont
  {Castelano}}, \ and\ \bibinfo {author} {\bibfnamefont {A.~O.}\ \bibnamefont
  {Caldeira}},\ }\href {http://stacks.iop.org/1367-2630/12/i=7/a=073009}
  {\bibfield  {journal} {\bibinfo  {journal} {New Journal of Physics}\ }\textbf
  {\bibinfo {volume} {12}},\ \bibinfo {pages} {073009} (\bibinfo {year}
  {2010})}\BibitemShut {NoStop}%
\bibitem [{\citenamefont {Poux}\ \emph {et~al.}(2014)\citenamefont {Poux},
  \citenamefont {Ara{\'u}jo}, \citenamefont {Filgueiras},\ and\ \citenamefont
  {Moraes}}]{poux2014}%
  \BibitemOpen
  \bibfield  {author} {\bibinfo {author} {\bibfnamefont {A.}~\bibnamefont
  {Poux}}, \bibinfo {author} {\bibfnamefont {L.~R.~S.}\ \bibnamefont
  {Ara{\'u}jo}}, \bibinfo {author} {\bibfnamefont {C.}~\bibnamefont
  {Filgueiras}}, \ and\ \bibinfo {author} {\bibfnamefont {F.}~\bibnamefont
  {Moraes}},\ }\href {\doibase 10.1140/epjp/i2014-14100-9} {\bibfield
  {journal} {\bibinfo  {journal} {The European Physical Journal Plus}\ }\textbf
  {\bibinfo {volume} {129}},\ \bibinfo {pages} {100} (\bibinfo {year}
  {2014})}\BibitemShut {NoStop}%
\bibitem [{\citenamefont {Abramowitz}\ and\ \citenamefont
  {Stegun}(1972)}]{abramo}%
  \BibitemOpen
  \bibfield  {author} {\bibinfo {author} {\bibfnamefont {M.}~\bibnamefont
  {Abramowitz}}\ and\ \bibinfo {author} {\bibfnamefont {I.~A.}\ \bibnamefont
  {Stegun}},\ }\href@noop {} {\emph {\bibinfo {title} {Handbook of Mathematical
  Functions}}}\ (\bibinfo  {publisher} {Dover Publications, New York},\
  \bibinfo {year} {1972})\BibitemShut {NoStop}%
\bibitem [{\citenamefont {Ivlev}(2017)}]{lasersingular}%
  \BibitemOpen
  \bibfield  {author} {\bibinfo {author} {\bibfnamefont {B.~I.}\ \bibnamefont
  {Ivlev}},\ }\href {\doibase 10.1139/cjp-2016-0830} {\bibfield  {journal}
  {\bibinfo  {journal} {Canadian Journal of Physics}\ }\textbf {\bibinfo
  {volume} {95}},\ \bibinfo {pages} {514} (\bibinfo {year} {2017})}\BibitemShut
  {NoStop}%
\bibitem [{\citenamefont {Kowalski}\ and\ \citenamefont
  {Rembieliński}(2013)}]{kowalski}%
  \BibitemOpen
  \bibfield  {author} {\bibinfo {author} {\bibfnamefont {K.}~\bibnamefont
  {Kowalski}}\ and\ \bibinfo {author} {\bibfnamefont {J.}~\bibnamefont
  {Rembieliński}},\ }\href {\doibase
  http://dx.doi.org/10.1016/j.aop.2012.10.003} {\bibfield  {journal} {\bibinfo
  {journal} {Annals of Physics}\ }\textbf {\bibinfo {volume} {329}},\ \bibinfo
  {pages} {146 } (\bibinfo {year} {2013})}\BibitemShut {NoStop}%
\bibitem [{\citenamefont {Streda}(1982)}]{Streda}%
  \BibitemOpen
  \bibfield  {author} {\bibinfo {author} {\bibfnamefont {P.}~\bibnamefont
  {Streda}},\ }\href {http://stacks.iop.org/0022-3719/15/i=22/a=005} {\bibfield
   {journal} {\bibinfo  {journal} {Journal of Physics C: Solid State Physics}\
  }\textbf {\bibinfo {volume} {15}},\ \bibinfo {pages} {L717} (\bibinfo {year}
  {1982})}\BibitemShut {NoStop}%
\bibitem [{\citenamefont {Gusynin}\ and\ \citenamefont
  {Sharapov}(2006)}]{graphenecondutivity}%
  \BibitemOpen
  \bibfield  {author} {\bibinfo {author} {\bibfnamefont {V.~P.}\ \bibnamefont
  {Gusynin}}\ and\ \bibinfo {author} {\bibfnamefont {S.~G.}\ \bibnamefont
  {Sharapov}},\ }\href {\doibase 10.1103/PhysRevB.73.245411} {\bibfield
  {journal} {\bibinfo  {journal} {Phys. Rev. B}\ }\textbf {\bibinfo {volume}
  {73}},\ \bibinfo {pages} {245411} (\bibinfo {year} {2006})}\BibitemShut
  {NoStop}%
\end{thebibliography}%

\end{document}